\documentclass[aps,pre,superscriptaddress,citeautoscript, amsmath,amssymb,twocolumn]{revtex4-1}	
\usepackage{graphicx}
\usepackage{amsmath}
\usepackage{hyperref}
\usepackage{color}
\usepackage{bigints}
\usepackage{mathrsfs}
\usepackage{algorithm}
\usepackage{algpseudocode}
\usepackage{multirow}
\usepackage[utf8]{inputenc}

\definecolor{darkblue}{RGB}{8,81,156}
\hypersetup{colorlinks,breaklinks,
            linkcolor=darkblue,urlcolor=darkblue,
           anchorcolor=darkblue,citecolor=darkblue}

\date{\today}

\allowdisplaybreaks

    \definecolor{dark-purple}{RGB}{118,42,131}
    \definecolor{dark-green}{RGB}{27,120,55}
    \definecolor{light-purple}{RGB}{231,212,232}
    \definecolor{LIGHT-PURPLE}{RGB}{194,165,207}
    \definecolor{light-green}{RGB}{168,216,183}
    \definecolor{gray}{RGB}{186,186,186}
    \definecolor{super-dark-green}{RGB}{0,69,41}
    \definecolor{super-dark-purple}{RGB}{63,0,125}
    \definecolor{super-dark-blue}{RGB}{8,48,107}
    \definecolor{super-dark-red}{RGB}{165,0,38}
    
    \definecolor{super-dark-purple}{RGB}{64,0,75}
    \definecolor{super-dark-green}{RGB}{0,68,27}

\usepackage{tabularx,booktabs} 

\usepackage{array}
\usepackage{longtable}
\newcolumntype{L}[1]{>{\raggedright\let\newline\\\arraybackslash\hspace{0pt}}p{#1}}
\newcolumntype{C}[1]{>{\centering\let\newline\\\arraybackslash\hspace{0pt}}m{#1}}
\newcolumntype{R}[1]{>{\raggedleft\let\newline\\\arraybackslash\hspace{0pt}}m{#1}}

\setcitestyle{super}

\usepackage{atveryend}
\makeatletter
\let\origcitation\citation
\AtEndDocument{\def\mycites{\@gobble}%
  \def\citation#1{\g@addto@macro\mycites{,#1}\origcitation{#1}}}
\AtVeryEndDocument{\typeout{***^^JCited keys: \mycites^^J***}}
\makeatother

\makeatletter
\let\origcitation\citation
\AtEndDocument{\def\mycites{}%
  \def\citation#1{\g@addto@macro\mycites{#1^^J,}\origcitation{#1}}}
\AtVeryEndDocument{\newwrite\citeout\immediate\openout\citeout=\jobname.cit
  \immediate\write\citeout{\mycites}\immediate\closeout\citeout}
\makeatother

\begin{document}

\title{How to Quantify and Avoid Finite Size Effects in Computational Studies of Crystal Nucleation: The Case of Heterogeneous Ice Nucleation}

\author{Sarwar Hussain}
\email{sarwar.hussain@yale.edu}
\affiliation{Department of Chemical and Environmental Engineering, Yale University, New Haven, CT  06520}

\author{Amir Haji-Akbari}
\email{amir.hajiakbaribalou@yale.edu}
\affiliation{Department of Chemical and Environmental Engineering, Yale University, New Haven, CT  06520}

\begin{abstract}
\noindent
Computational studies of crystal nucleation can be impacted by finite size effects, primarily due to unphysical interactions between crystalline nuclei and their periodic images. It is, however, not always feasible to systematically investigate the sensitivity of nucleation kinetics and mechanism to system size due to large computational costs of nucleation studies.  Here, we use jumpy forward flux sampling to accurately compute the rates of heterogeneous ice nucleation in the vicinity of square-shaped model structureless ice nucleating particles (INPs) of different sizes, and identify three distinct regimes for the dependence of rate on the INP dimension, $L$. For small INPs, the rate is a strong function of $L$ due to artificial spanning of critical nuclei across the periodic boundary. Intermediate-sized INPs, however, give rise to the emergence of non-spanning 'proximal` nuclei that are close enough to their periodic images to fully structure the intermediary liquid. While such proximity can facilitate nucleation, its effect is offset by the {\color{black}higher density} of the intermediary liquid, leading to artificially small nucleation rates overall. The critical nuclei formed at large INPs are neither spanning nor proximal. Yet, the rate is a weak function of $L$, with its logarithm scaling linearly with $1/L$. The key heuristic emerging from these observations is that finite size effects will be minimal if critical nuclei are neither spanning nor proximal, and if the {\color{black}intermediary liquid has a region that is structurally indistinguishable from the supercooled liquid under the same conditions.}
\end{abstract}
\maketitle

\section{Introduction}
\label{section:intro}

\noindent
The main premise of molecular simulations is to use the information obtained from simulating finite-sized systems to predict their behavior in the thermodynamic limit. The accuracy of such predictions, however, can depend strongly on the size of the simulated system, as estimates of thermodynamic,\cite{BinderFerroelectrics1987, MonJChemPhys1992, HorbachPhysRevE1996, AguadoJChemPhys2001, OreaJCP2005, MastnyJChemPhys2007, BiscayJCP2009, BurtJPhysChemC2016} structural,\cite{SalacusePhysRevE1996} and transport\cite{JamaliJChemTheoryComput2018} properties and nucleation rates~\cite{MichaelidesChemRev2016} in small systems can deviate  from those in the thermodynamic limit in a statistically significant manner. Such a dependence on system size is typically referred to as \emph{finite size effects}, which can be fairly strong for very small systems, while being mostly unnoticeable for larger systems. Therefore, finite size effects can, in principle, be mitigated by simulating very large systems, a task that is only computationally feasible for simple model systems.\cite{SwopePhysRevB1990, EnglishPhysRevE2015} Fortunately, this is not always necessary as similar conclusions can usually be obtained from simulating "sufficiently large`` computationally tractable systems.\cite{SwopePhysRevB1990, WedekindJCP2006} What constitutes "sufficiently large``, however, is subject to the property that is being estimated or the process that is being studied. For instance, a system comprised of several hundred molecules is usually large enough for accurately estimating thermodynamic, structural and transport properties of  liquids,\cite{MandellJStatPhysics1976, BinderFerroelectrics1987, MonJChemPhys1992, HorbachPhysRevE1996, SalacusePhysRevE1996, JamaliJChemTheoryComput2018} but might be too small for studying collective phenomena such as cavitation,\cite{MeadleyJCP2012} condensation\cite{PerezJCP2011} and crystal nucleation.\cite{HoneycuttChemPysLett1984, HoneycuttJPC1986, SwopePhysRevB1990} It is therefore critical to develop heuristics for determining what qualifies as "sufficiently large`` for studying such collective phenomena, in order to ensure the accuracy and reliability of the conducted simulations.  

One such collective phenomenon that has been extensively studied using molecular simulations is crystal nucleation. As such, understanding the role of finite size effects on the thermodynamics and kinetics of crystal nucleation has been a topic of interest for decades.\cite{MichaelidesChemRev2016} Nucleation is a process in which a sufficiently large nucleus of the new phase forms within the old metastable phase, and is usually the rate-limiting step of a first-order phase transition when the underlying thermodynamic driving force is small.\cite{FalahatiSoftMatter2019}  Finite size effects in nucleation primarily arise due to periodic boundary conditions, which can result in an unphysical confinement of the metastable phase between the nucleus and its periodic images,\cite{MandellJStatPhysics1976, HoneycuttChemPysLett1984, HoneycuttJPC1986, PengJCP2010} or the formation of nuclei that span across the periodic boundary.\cite{MeadleyJCP2012, StattPRL2015} In the case of crystal nucleation, the effect of periodic boundaries might be stronger due the extension of the diffuse crystal-liquid interface beyond the nucleus.\cite{SwopePhysRevB1990} However, finite size effects can also  arise due to other factors such as solute depletion in multi-component systems,\cite{SalvalaglioPNAS2015, LiuMolPhys2018} or peculiarities of the employed ensemble.\cite{WedekindJCP2006, PerezJCP2011}  These effects can collectively lead to unphysical nucleation rates in both homogeneous and heterogeneous nucleation, and has also been found to impact crystal growth.\cite{BurkeJCP1988, OMalleyPRL2003, StreitzPRL2006} Indeed, the findings of several high-profile computational studies of nucleation are believed to have been strongly impacted by finite size effects. For instance, Matsumoto~\emph{et al.}'s  observation\cite{Matsumoto2002} of homogeneous ice nucleation in a system of 512 water molecules represented using the fully atomistic TIP4P\cite{JorgensenJChemPhys1983} model has never been reproduced in larger systems, and was later shown  to be an artifact of strong finite size effects.\cite{SanzJACS2013} Earlier computational studies of surface freezing-- or surface-induced homogeneous ice nucleation\cite{TabazadehPNAS2002, HajiAkbariJCP2017}-- by Vrbka and Jungwirth\cite{JungwithJPCB2006, JungwirthJMolLiq2007} and Pluharova~\emph{et al.}\cite{JungwirthJPhysChemC2010} are also believed to be impacted by finite size effects.\cite{HajiAkbariPNAS2017}

Early efforts to characterize finite size effects in crystal nucleation focused on homogeneous nucleation in the simple Lennard-Jones (LJ) liquid.\cite{LJProcRSoc1924} For instance, Honeycutt and Andersen~\cite{HoneycuttChemPysLett1984, HoneycuttJPC1986} simulated systems of up to 1,500 LJ particles at a reduced density of 0.95 and a reduced temperature of 0.45 and concluded that the occurrence of the "catastrophic crystal growth`` observed in earlier simulations of the deeply supercooled LJ liquid is not due to the emergence of a critical nucleus and is instead an artifact of periodic boundaries. Indeed, they later demonstrated that critical nuclei form way earlier than the catastrophic growth, but their average sizes and  the time needed for their formation both tend to increase with system size, pointing to strong finite size effects even prior to catastrophic growth.\cite{HoneycuttJPC1986} Later, Swope and Andersen\cite{SwopePhysRevB1990} conducted large-scale MD simulations of 15,000 and $10^6$ LJ particles under similar conditions, and observed that the properties of the 15,000-particle system were similar to the average properties of the 64 subsystems within the million-particle simulation box. They therefore concluded that  the 15,000-particle system is large enough to be devoid of finite size effects.  A similar conclusion was reached by Huitema \emph{et al.},\cite{HuitemaPhysRevB2000} who examined homogeneous nucleation in systems with as many as 10,000 particles. According to these studies, avoiding finite size effects requires simulating systems that are at least three orders of magnitude larger than the characteristic critical nucleus size, a very stringent requirement that can only be satisfied for simple model systems. Consequently, such large-scale simulations of finite size effects in other systems are rare.\cite{EnglishPhysRevE2015} This heuristic is, however, based on observations in the high-rate regime, i.e.,~where nucleation can occur during a computationally tractable MD trajectory, and therefore the liquid structure is pre-disposed to freezing. It is therefore plausible to expect finite size effects to be weaker in the low-rate regime in which nucleation events are spatially isolated. Moreover, this heuristic can only apply to homogeneous nucleation at best, as the dependence of rate on system size might be completely different for heterogeneous nucleation. 

Unfortunately, many of these questions are yet to be investigated in a systematic manner. Consequently, there are no rigorous guidelines or heuristics for avoiding finite size effects in crystal nucleation studies, and different authors have resorted to different \emph{ad hoc} approaches to avoid finite size artifacts. While some have tested the robustness of their findings by repeating their simulations in computationally tractable larger systems,\cite{CoxFaraday2013, CoxJCP2015, CoxIIJCP2015} others have argued that finite size effects will be absent if the average distance between the critical nucleus and its periodic images is larger than half the box dimensions.\cite{SossoJPhysChemLett2016, LupiJCP2016} This latter heuristic usually translates to critical nuclei that are at least an order of magnitude smaller than the system size in homogeneous nucleation, a heuristic that is sometimes referred to as the "10\% rule``. There are, however, reasons to doubt the adequacy of these approaches. The former approach can only be conclusive if rate calculations are conducted for a wide range of system sizes, an undertaking that is usually not feasible. The size and the distance thresholds invoked in the latter approach, on the other hand, are not based on any rigorous analysis, and can be fairly sensitive to the particulars of the algorithm utilized for detecting crystalline nuclei. 

Here, we attempt to address some of these questions by systematically investigating how the rate and mechanism of heterogeneous ice nucleation within supercooled supported water nanofilms in the vicinity of a model structureless ice nucleating particle (INP) is affected by system size.  Due to the surface-dominated nature of heterogeneous nucleation, the relevant "system size`` is  the dimensions of the INP. We use our recently developed jumpy forward flux sampling (jFFS)~\cite{HajiAkbariJCP2018} algorithm to compute nucleation rates for 16 different system sizes, comprising between 1,600 and 50,176 water molecules. We identify three distinct regimes for the dependence of rate and mechanism on system size, and identify the critical nuclei characteristics that signify each regime. Based on our observations, we devise a rigorous set of heuristics for assessing whether a particular nucleation rate calculation is impacted by finite size effects. Moreover, we provide a scaling approach for estimating the rate in the thermodynamic limit for system sizes where finite size effects are minimal but the rate is still a weak function of system size.

\begin{figure*}
\centering
\includegraphics[width=0.6\textwidth]{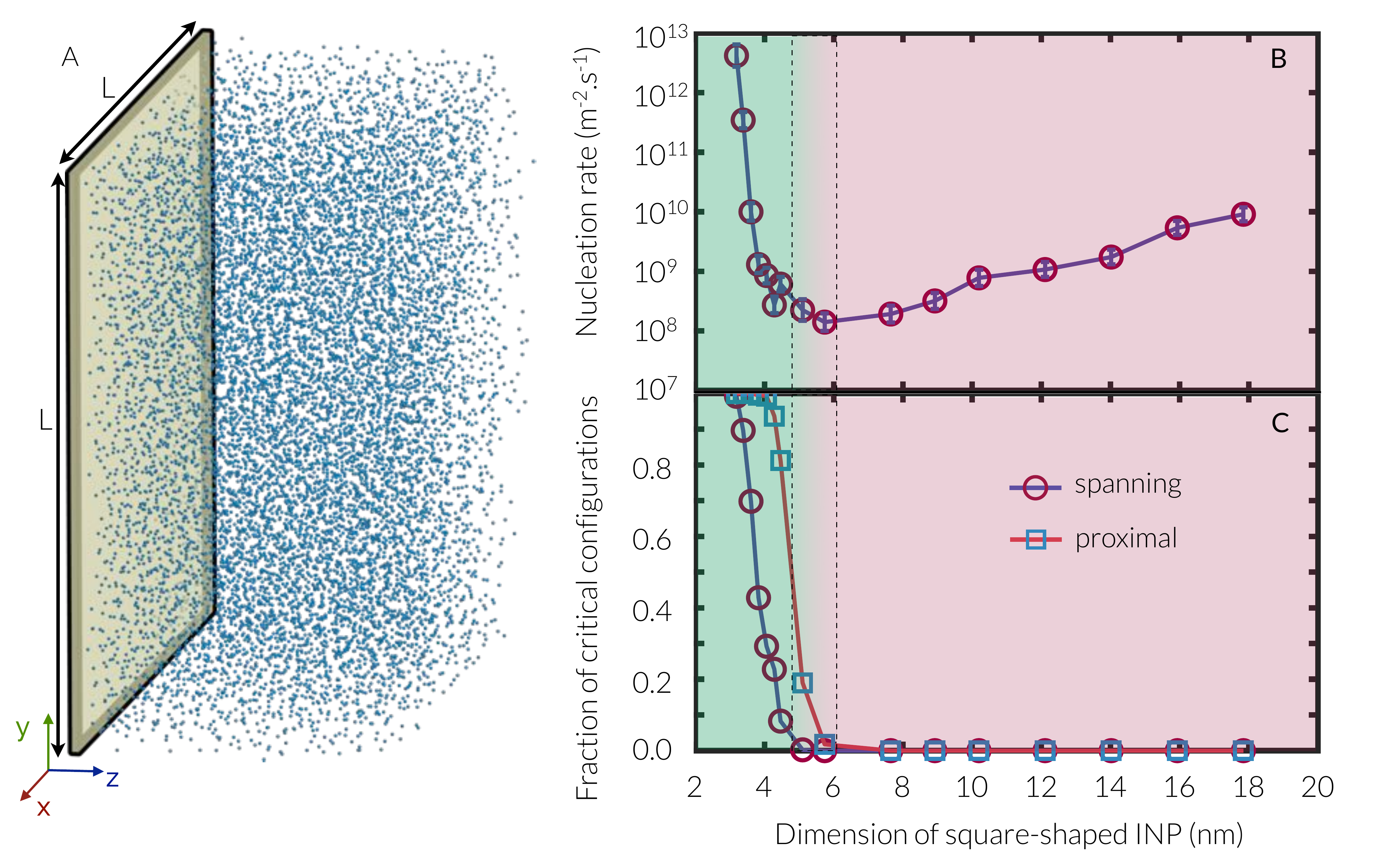} 
\caption{{\color{black}(A) Schematic representation of a supported film of supercooled water (dark blue) in the vicinity of a structureless INP (light green)} (B) The dependence of heterogeneous nucleation rate on $L$, the dimension of the square-shaped INP. (C) The fraction of critical nuclei that are  spanning (circles) and  proximal (squares). Error bars correspond to 95\% confidence intervals and are smaller than the symbols.}
\label{fig:schematics-rates}
\end{figure*}

\section{Methods}
\label{section:methods}

\subsection{System Description and Molecular Dynamics Simulations} 

\noindent 
We consider heterogeneous nucleation in supported nanofilms of supercooled water in the vicinity of a model  structureless INP at a temperature of 235~K. Unlike some earlier studies\cite{BiJPhysChemC2016, LupiJCP2016} of heterogeneous ice nucleation in which the liquid film is sandwiched between the INP and its periodic image, the films considered in this work only touch the INP on one side, and are exposed to vacuum at the other interface {\color{black}(Fig.~\ref{fig:schematics-rates}A)}. Despite being more expensive computationally, we believe that our setup constitutes a more faithful representation of  heterogeneous nucleation in nature, where isolated INPs are in contact with a sea of the supercooled liquid. The water nanofilms considered in this work are  approximately 4.8 nm thick, which makes it extremely unlikely for the free interface to impact the kinetics and mechanism of nucleation in a meaningful manner.\cite{Hussain2020}  

We model water molecules using the monoatomic water (mW) potential,\cite{MolineroJPCB2009} a popular coarse-grained model of water obtained via re-parameterizing the Stillinger-Weber (SW) potential originally developed for modeling Group IV elements such as carbon and silicon.\cite{StillingerPRB1985} The model INP is square-shaped and is located in the  $xy$ plane. It interacts with water molecules  via the Lennard-Jones (LJ) 9-3 potential~\cite{MagdaJCP1985} with  $\epsilon = 1.2~\text{kcal}\cdot\text{mol}^{-1}$ and $\sigma = 0.32~\text{nm}$ and with a cutoff of 0.8~nm. These parameters are chosen so that the underlying INP does not induce ice nucleation very strongly.  The utilized $\epsilon$ value, in particular, is the smallest $\epsilon$ for which heterogeneous nucleation is observed in unbiased 50-ns long MD simulations at 215~K. All MD simulations are performed in the canonical (NVT) ensemble using LAMMPS.\cite{PimptonLAMMPS1995} It is necessary to emphasize that the presence of a vapor-liquid interface in our simulation setup makes it possible for the liquid film to freely 'adjust` to density changes during nucleation. This is equivalent to simulating  nucleation under an effective zero pressure. Our calculations are therefore not susceptible to artifacts due to unphysical density fluctuations in bulk finite systems simulated in the NVT ensemble\cite{WedekindJCP2006}. Equations of motion are integrated using the velocity-Verlet algorithm with a time step of 5 fs, while temperature is controlled using the Nos\'e-Hoover~\cite{NoseMolPhys1984, HooverPhysRevA1985} thermostat with a time constant of 0.5 ps. {\color{black}
While conventional thermostats are incapable of accounting for the effect of latent heat of fusion, they  have been previously shown to perform reasonably well in predicting the nucleation rate, due to the non-Gaussian distribution of temperatures of pre- and post-critical nuclei.\cite{WedekindJChemPhys2007p} As such, we do not expect any of our findings to be strongly impacted by the employed thermostat. }

\begin{table}
\centering
\vspace{-6pt}
\caption{Heterogeneous ice nucleation rates computed at 235~K using jFFS. $N_p$ refers to the number of water molecules within each film. {\color{black}The} error bars {\color{black}in $\log_{10}\mathcal{J}$} correspond to 95\% confidence intervals {\color{black}while uncertainties in $N^*$ correspond to the range of $\lambda$'s for which $0.35\le p_c(\lambda)\le 0.65$}.}
\centering
\begin{tabular}{R{1.7cm}R{1.7cm}L{2.0cm}R{2.5cm}}
\hline\hline
$N_p$~~ & $L~[\text{nm}]$& ~~~~~~$N^*$ & $\log_{10}\mathcal{J}[\text{m}^{-2}\cdot\text{s}^{-1}]$\\  
\hline
1,600&3.1869& ~~~{\color{black}$145\pm9$} & $12.6243\pm0.1882$\\ 
2,304&3.3803& ~~~{\color{black}$153\pm10$} & $11.5403\pm0.1334$\\ 
2,304&3.6056& ~~~{\color{black}$178\pm10$} & $9.9990\pm0.1468$\\ 
2,304&3.8243& ~~~{\color{black}$196\pm10$} & $9.1136\pm0.1446$\\ 
3,136&4.0563& ~~~{\color{black}$230\pm13$} & $8.9292\pm0.1302$\\ 
3,136&4.2817& ~~~{\color{black}$248\pm14$} & {\color{black}$8.4352\pm0.1380$}\\ 
3,136&4.4617& ~~~{\color{black}$271\pm17$} & {\color{black}$8.7846\pm0.1218$}\\ 
4,096&5.0991&  ~~~{\color{black}$272\pm17$} & $8.3486\pm0.1802$\\ 
5,184&5.7365&  ~~~{\color{black}$276\pm17$} & $8.1461\pm0.1474$\\ 
9,216 &7.6487&  ~~~{\color{black}$280\pm18$} & {\color{black}$8.2822\pm0.1430$}\\ 
12,544 &8.9234&  ~~~{\color{black}$268\pm15$} & $8.5019\pm0.1336$\\ 
16,384&10.1983&  ~~~{\color{black}$271\pm17$} & $8.8893\pm0.1378$\\ 
23,104&12.1104&  ~~~{\color{black}$272\pm17$} & $9.0289\pm0.1246$\\ 
30,976&14.0226&  ~~~{\color{black}$274\pm18$} & $9.2391\pm0.1210$\\ 
40,000&15.9348 &  ~~~{\color{black}$272\pm18$} & $9.7301\pm0.1178$\\ 
50,176&17.8470&  ~~~{\color{black}$274\pm18$} & $9.9583\pm0.1160$ \\[1ex] 
\hline\hline
\end{tabular}
\vspace{-20pt}
\label{table:rates}
\end{table}

As mentioned above, we utilize $L$, the dimension of the structureless INP, as a proxy for system size. We consider a total of 16 different $L$'s, ranging from $3.19~\text{nm}$ to $17.85~\text{nm}$, and use the following procedure for initializing the water nanofilms. We first generate a properly sized slab of cubic ice comprised of $n\times n\times8$ unit cells. The oxygen-oxygen distance in each unit cell, $r_{OO}$ is adjusted so that the target $L$ is an integer multiple of $L_c=4r_{OO}/\sqrt3$, the unit cell dimension of cubic ice.\cite{HajiAkbariAFPSpringer2020} For most $L$'s, we use the value of $r_{OO}=0.276$~nm, while for a few smaller system sizes, $r_{OO}$ is slightly adjusted in order to fit an integer number of unit cells within the box. These include boxes with lateral dimensions of 3.38, 3.61, 4.06 and 4.28~nm, for which the $r_{OO}$ values of 0.24, 0.26, 0.25 and 0.26~nm are utilized, respectively. Each ice film is then melted at a temperature of 300~K. A minimum of 100 configurations are saved along the melted trajectory every 50~ps, which are then quenched to the target temperature of 235~K at a cooling rate of 7.69~$\text{ps}\cdot\text{K}^{-1}$. A list of all $L$'s as well as the number of water molecules within each film is given in Table~\ref{table:rates}.

\subsection{Rate Calculations}

\subsubsection{Order Parameter}
\noindent
We compute nucleation rates using our recently developed jFFS algorithm,\cite{HajiAkbariJCP2018} which is a generalized variant of the forward flux sampling algorithm (FFS)\cite{FrenkelFFS_JCP2006} that has been extensively utilized to study rare events.\cite{HussainJChemPhys2020} Similar to most other advanced sampling techniques, conducting an FFS calculation requires an order parameter, a mathematical function $\lambda:\mathcal{Q}\rightarrow\mathbb{R}$ that quantifies the progress of the corresponding rare event, in this case, heterogeneous ice nucleation. Here, $\mathcal{Q}$ is the configuration space that includes all the microscopic degrees of freedom of the corresponding system,~i.e.,~the positions of all water molecules. In this work, we use the number of molecules in the largest crystalline nucleus as the order parameter. Each molecule $i$ is classified as solid-like or liquid-like based on $q_6(i)$, the Steinhart bond order parameter,\cite{SteinhardtPRB1983} given by,
\begin{eqnarray}
q_l(i) &=& \frac{1}{N_b(i)} \sum_{j=1}^{N_b(i)} \frac{\textbf{q}_l(i)\cdot\textbf{q}_{l}^*(j)}{|\textbf{q}_l(i)||\textbf{q}_l(j)|}.
\end{eqnarray}
Here, $N_b(i)$ is the number of water molecules within a distance of $r_c=0.32~\text{nm}$  from $i$ {\color{black}and $r_c$ corresponds to the locus of the first valley of the radial distribution function. Therefore, the neighbors of each molecule are those that lie within its first hydration shell.} $\textbf{q}_l(i)$ is a $(2l+1)$-component complex-valued vector, and its components are given by,
\begin{eqnarray}
q_{lm} (i) = \frac{1}{N_b(i)} \sum_{j=1}^{N_b(i)} Y_{lm}(\theta_{ij},\phi_{ij}),~~-l\le m\le+l 
\end{eqnarray}
with $\theta_{ij}$ and $\phi_{ij}$  the polar and azimuthal angles corresponding to the separation vector $\textbf{r}_{ij}=\textbf{r}_j-\textbf{r}_i$, and $Y_{lm}(\cdot,\cdot)$'s, the spherical harmonic functions. Consistent with our earlier studies,\cite{HajiAkbariFilmMolinero2014, HajiAkbariPNAS2015, GianettiPCCP2016, HajiAkbariPNAS2017, HajiAkbariJCP2018} we classify molecules with $q_6\ge0.5$ as solid-like, cluster the solid-like molecules that are within a distance of $r_c$ into crystalline nuclei, and  apply the chain exclusion algorithm of Reinhardt~\emph{et al.}~\cite{VegaJCP2012} 

\subsubsection{jFFS Calculations}\label{section:jFFS}

\noindent
With $\lambda(\cdot)$ at hand, the transition region between the supercooled liquid basin $A := \{x\in\mathcal{Q}: \lambda(x) < \lambda_A\}$ and the crystalline basin $B := \{x\in\mathcal{Q}: \lambda(x)\ge \lambda_B\}$ is partitioned into $N$ non-overlapping regions using $N$ milestones $\lambda_A<\lambda_0<\lambda_1\cdots\lambda_N=\lambda_B$, which are level sets of $\lambda(\cdot)$. As demonstrated earlier,\cite{HajiAkbariJCP2018} our utilized $\lambda(\cdot)$ is jumpy as it undergoes high-frequency high-amplitude fluctuations along an MD trajectory. We therefore need to use jFFS in order to accurately capture the potentially subtle changes in rate upon changing $L$. In order to minimize the number of FFS iterations, we follow the approach described in Section III B 1 of Ref.~\citenum{HajiAkbariJCP2018}. For each $L$, we first conduct conventional MD trajectories within $A$ with a combined duration of at least 0.5~$\mu$s, and monitor for first crossings of $\lambda_0$. The configurations corresponding to such crossings $\mathcal{C}_0=\{x_1^{(0)},x_2^{(0)},\cdots,x_{N_0}^{(0)}\}$ are stored, and $\Phi_0$, the flux of trajectories that leave $A$ and cross $\lambda_0$, is given by,
\begin{eqnarray}
\Phi_0 &=& \frac{N_0}{\mathcal{T}L^2},
\end{eqnarray}
with $\mathcal{T}$ the combined duration of the MD trajectories and $L^2$ the surface area of the underlying INP. The next step is to compute the probability that a trajectory initiated from $\mathcal{C}_0$ reaches $B$ by recursively computing the transition probabilities between successive milestones. In the simplified scheme of jFFS utilized here, the intermediate milestones are chosen on the fly. In particular, $\lambda_k$ is chosen so that it is beyond $\lambda_{k-1,\max}=\max_{x\in\mathcal{C}_{k-1}} \lambda(x)$, wherein $\mathcal{C}_{k-1}$ contains all the configurations obtained upon a first crossing of $\lambda_{k-1}$. After choosing the next target milestone $\lambda_k$, a large number of trial trajectories are initiated from the configurations in $\mathcal{C}_{k-1}$ with their  momenta sampled from the Maxwell-Boltzmann distribution, and each trajectory is terminated upon crossing $\lambda_k$ or returning to $A$. The transition probability $P(\lambda_k|\lambda_{k-1})$ is then computed as the fraction of trial trajectories that cross $\lambda_k$.  Note that $\lambda_B$ is not known \emph{a priori}. Instead, the calculation is terminated when $P(\lambda_k|\lambda_{k-1})$ is statistically indistinguishable from unity. The nucleation rate $\mathcal{J}$ is then computed from,
\begin{eqnarray}
\mathcal{J} &=& \Phi_0\prod_{k=1}^N P(\lambda_k|\lambda_{k-1}),
\end{eqnarray}
The statistical uncertainty of the computed $\mathcal{J}$ is estimated using the approach described in Ref.~\citenum{FrenkelFFS_JCP2006}. All reported error bars correspond to 95\% confidence intervals, i.e.,~twice the standard errors obtained from this approach. {\color{black}Further technical details of the jFFS calculations, including crossing statistics,  FFS milestones and transition probabilities, and the average length of successful and unsuccessful trajectories in each iteration are given in the Supplementary Information. }

It has been previously demonstrated in numerous  studies~\cite{LupiJCP2016, LupiNature2017} that $\lambda(\cdot)$ is a good reaction coordinate for crystal nucleation. This implies that the critical nucleus size can be accurately determined from the committor probability given by,
\begin{eqnarray}
p_c(\lambda_k) &=& \prod_{q=k+1}^N P(\lambda_q|\lambda_{q-1}),\label{eq:p-c}
\end{eqnarray}
More precisely, $N^*$, the critical nucleus size, is estimated by fitting $p_c(\lambda)$ to the following expression,
\begin{eqnarray}
p_c(\lambda) &=& \dfrac12\bigg\{1+\text{erf}\left[a(\lambda-N^*)\right]\bigg\},\label{eq:pc-erf-fit}
\end{eqnarray}
with the reported error bar{\color{black}s} corresponding to the range of $\lambda$'s for which $0.35\le p_c(\lambda)\le 0.65$. This choice is motivated by the fact that in an FFS rate calculation, only a handful of FFS milestones are close the transition region,~i.e.,~the range of $\lambda$'s for which the committor probability is close to 50\%. As a result, the attained estimates of $p_c(\lambda)$-- and $N^*$-- can be potentially sensitive to milestone placement in FFS. Therefore, other measures of uncertainty, such as the confidence interval of the fit parameter of Eq.~(\ref{eq:pc-erf-fit}), might underestimate the true uncertainty in $N^*$, while the approach outlined here is more robust to factors such as milestone placement. It is necessary to emphasize that none of these issues impact the overall rate estimates, which are known to be insensitive to the number and location of milestones\cite{HajiAkbariJCP2018}.

 In probing the properties of critical configurations, we identify $\lambda_{k^*}$, the closest milestone to $N^*$, and denote all $x\in\mathcal{C}_{k^*}$ as critical. For most systems, this results in configurations with $32\%\le p_c(\lambda(x))\le 68\%$. In {\color{black}three} systems,~i.e.,~$L={\color{black}7.65,}~12.11$ and $14.02$~nm, $N^*$ is not sufficiently close to any of the milestones, so we choose the configurations obtained from first crossings of the two closest milestones, which results in committor probabilities no smaller than 16\% and no larger than 78\%.

\subsubsection{Analysis of Nucleation Mechanism and Identification of Bottlenecks}

\noindent
Since a successful nucleation pathway in jFFS is generated sequentially by concatenating the trial trajectories between successive milestones, we can trace back the ancestry of any configuration in $\mathcal{C}_N$, and identify the \emph{surviving configurations} in earlier milestones, i.e.,~those with some progeny at $\lambda_B$. We denote the surviving subset of $\mathcal{C}_k$ as $\mathcal{C}_k^s$.  By comparing the properties of $\mathcal{C}_k^s$ and $\mathcal{C}_k$, we can identify the important features that play a key role in successful nucleation. More specifically, for a given mechanical observable $\mu:\mathcal{Q}\rightarrow\mathbb{R}$, one can compute $\mu_s(\lambda_k)=\langle \mu(x)\rangle_{x\in\mathcal{C}_k^s}$ and $\mu_a(\lambda_k)=\langle \mu(x)\rangle_{x\in\mathcal{C}_k}$. We denote the milestone at which $\mu_s(\lambda)$ and $\mu_a(\lambda)$ cross each other a \emph{bottleneck} for $\mu(\cdot)$, and the corresponding average $\mu_b$.  As demonstrated in our prior applications of this approach,\cite{HajiAkbariPNAS2015, HajiAkbariPNAS2017, AltabetPNAS2017, HajiAkbariJCP2018} the existence of a noticeable bottleneck indicates that $\mu(\cdot)$ is {\color{black}likely} an important feature (orthogonal to the order parameter) that determines the likelihood of an early-milestone configuration to succeed in having progeny at $\lambda_B$.

{\color{black}
\subsection{Characterizing the Proximity of Crystalline Nuclei to Their Periodic Images}
\label{section:methods:spanning}

\noindent
Here, we describe the procedure for quantifying the extent by which a crystalline nucleus is impacted by its closest periodic images. Let  $\mathcal{N}=\{\textbf{r}_1,\textbf{r}_2,\cdots,\textbf{r}_m\}$ be a crystalline nucleus comprised of $m$ molecules within a box with side vectors $\textbf{b}_x, \textbf{b}_y$ and $\textbf{b}_z$. The periodic images of $\mathcal{N}$ can be obtained by adding integer linear combinations of $\textbf{b}_x, \textbf{b}_y$ and $\textbf{b}_z$ to $\textbf{r}_i$'s. More precisely, the $n_xn_yn_z$-th periodic image of $\mathcal{N}$ will be given by,
$$
\mathcal{N}_{n_xn_yn_z}=\{\textbf{r}_i+n_x\textbf{b}_x+n_y\textbf{b}_y+n_z\textbf{b}_z\}_{i=1}^m,~~~n_x,n_y,n_z\in\mathbb{Z}
$$
We are particularly interested in the closest periodic images of $\mathcal{N}$, and thus in a subset of displacement vectors $\mathcal{P}$ that produce such closest images. For heterogeneous nucleation on the square-shaped INPs considered in this work, $\mathcal{P}$ is given by $\mathcal{P}=\{(\pm L, 0, 0), (0,\pm L,0)\}$. It must, however, be noted that $\mathcal{P}$, in general, will depend on the shape of the simulation box, and on whether nucleation is homogeneous or heterogeneous. For instance, for homogeneous nucleation within a cubic box of side $L$, $\mathcal{P}$ will be given by $\{(\pm L,0,0), (0,\pm L, 0), (0, 0, \pm L)\}$.

\begin{figure*}
\centering
\includegraphics[width=.65\textwidth]{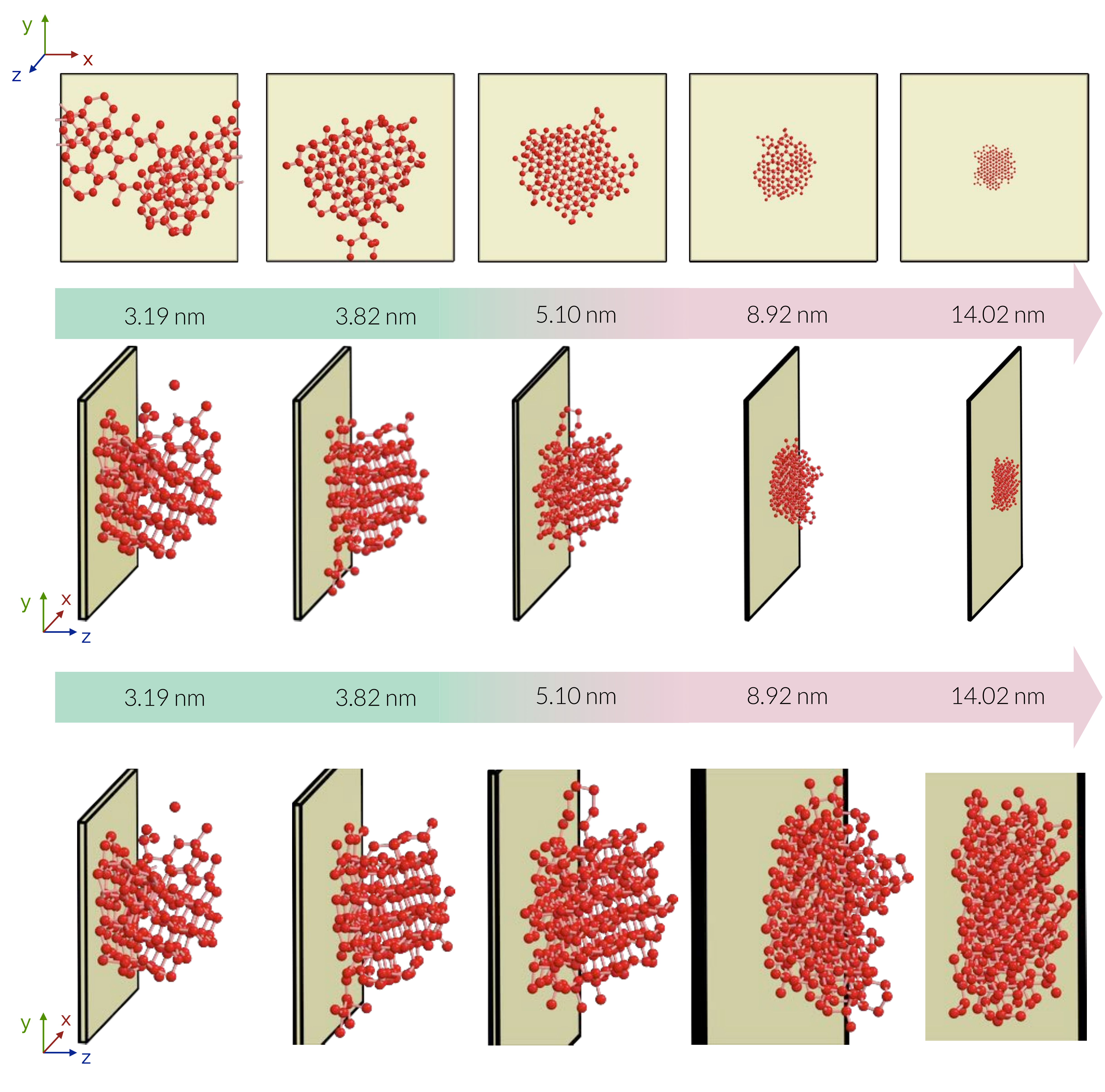}
\caption{\label{Fig:critical}{\color{black} Representative critical crystalline nuclei from nucleation on five INPs of different sizes depicted from two different angles, with water molecules belonging to the nucleus and the structureless INP depicted in dark red and light green, respectively. The color code utilized for distinguishing different regimes  is identical to that of Fig.~\ref{fig:schematics-rates}. The configurations in the third row are zoomed-in versions of those in the second row.}}
\end{figure*}

The most extreme manifestation of finite size effects is when $\mathcal{N}$ is \emph{spanning},~i.e.,~when it is connected to some-- or all-- of its periodic images. In order to determine whether  $\mathcal{N}$ is spanning, we label each molecule within $\mathcal{N}$ with a set of (up to three) integers $\pmb\xi=(\xi_{x},\xi_{y},\xi_{z})$ that describe the side of the boundary that the molecule resides with respect to a reference molecule. For heterogeneous nucleation, spanning is only possible across two dimensions so only two indices are needed. At the beginning of the analysis, we set all $\pmb\xi_i$'s to zero except for that of an arbitrary reference molecule $j$ within $\mathcal{N}$, which is set to $\pmb1$. ($\pmb1$ stands for a vector of all 1's.)  We then recursively loop through the nearest neighbors of each molecule in $\mathcal{N}$  (starting with $j$) and set the $\pmb\xi$'s of the unvisited molecules depending on whether reaching that neighbor requires crossing the boundary. More specifically, if $q$ is an unvisited neighbor of $p$ and is located at the same side of the periodic boundary as $p$, $\pmb\xi_q$ is set to $\pmb\xi_p$. Otherwise, the index corresponding to the boundary that needs to be crossed is multiplied by $-1$. For instance, if reaching $q$ requires crossing the $x$ boundary, $\xi_{q,x}$ is set to $-\xi_{p,x}$.  
In order for $\mathcal{N}$ to be spanning, it needs to contain two neighboring molecules with differing $\pmb\xi$'s, but located at the same side of the periodic boundaries.

\begin{figure}
\centering
\includegraphics[width=0.48\textwidth]{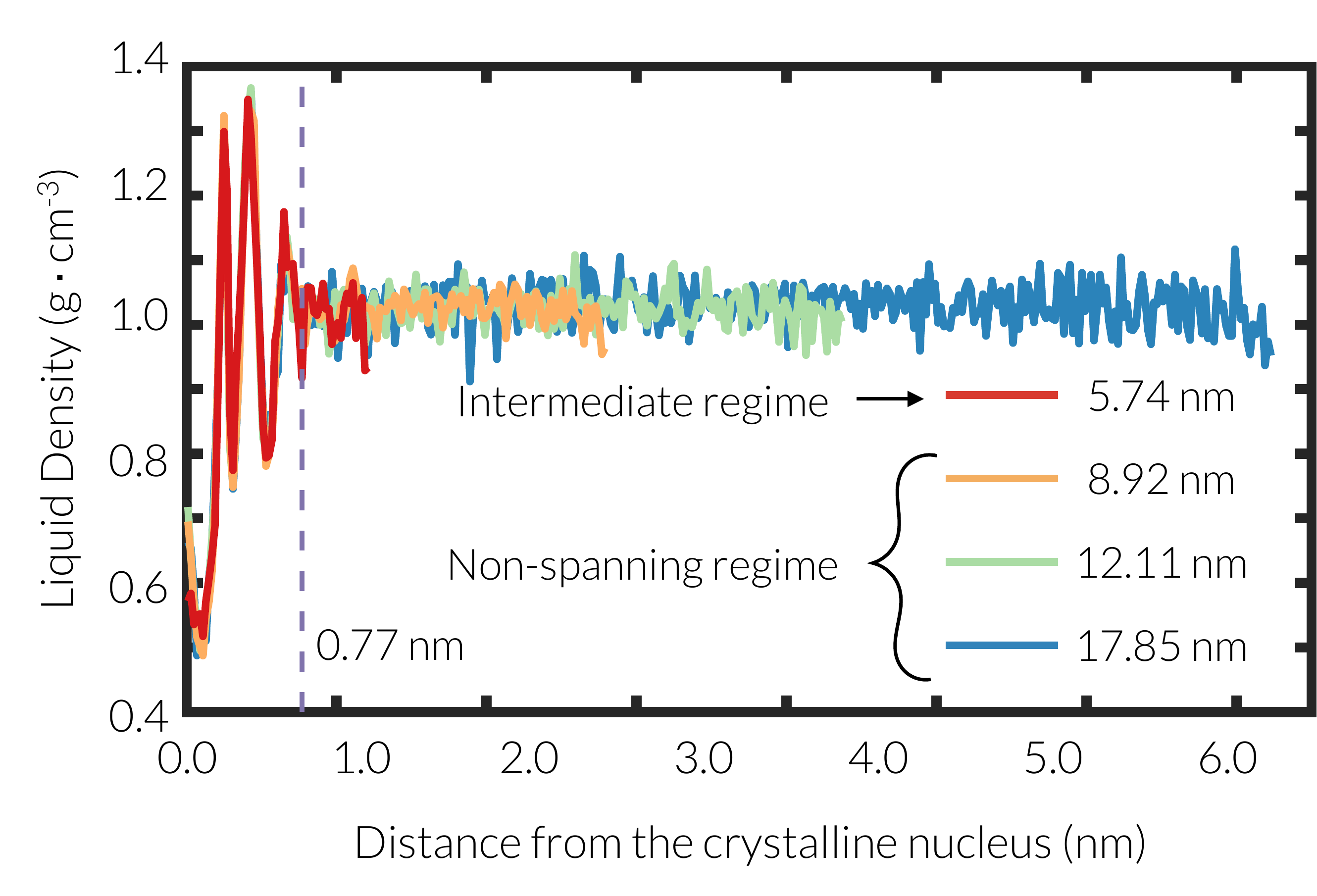} 
\caption{$\rho_{\textbf{u}}(r)$, the inter-image liquid density as a function of the distance from the critical cluster for a few representative values of $L$.}
\label{fig:interimage-density}
\end{figure}

In order to explore the structure of the liquid that is confined between a non-spanning nucleus $\mathcal{N}$ and its closest periodic image, we determine $\textbf{u}$, the shortest vector connecting $\mathcal{N}$ to its closest periodic image, and calculate the liquid density profile along $\textbf{u}$. In order to determine $\textbf{u}$, we first identify the range of $x, y$ (or $z$) values for all $\textbf{r}_i\in\mathcal{N}$ that have identical $\pmb\xi_i$'s, and use the attained intervals to displace all  molecules in the system so that (the shifted) $\mathcal{N}$ is not partitioned by the periodic boundaries. Note that relabeling such a shifted nucleus will yield $\pmb\xi=\pmb1$ for all its constituent molecules.  {\color{black}It must also be emphasized that such shifting can be conducted for non-spanning nuclei only, since spanning nuclei will always cross some periodic boundaries. Therefore, the labeling procedure described above is a pre-requisite for such shifting as it enables us to confirm that a particular nucleus is non-spanning.} The inter-image vector $\textbf{u}$ can then be determined as,
\begin{eqnarray}
[\tilde{i},\tilde{j},\tilde{\textbf{p}}] &=& \text{argmin~}_{i,j\le m,\textbf{p}\in\mathcal{P}} |\textbf{r}_i - \textbf{r}_j^{\textbf{p}}|, \label{eq:argmin-uvec}\\
\textbf{u} &=& \textbf{r}_{\tilde{j}}^{\tilde{\textbf{p}}} - \textbf{r}_{\tilde{i}}, \label{eq:uvec}
\end{eqnarray}
We call $\textbf{u}$ the \emph{inter-image vector} and the supercooled liquid that resides along it the \emph{inter-image liquid}. Our next step is to compute $\rho_{\textbf{u}}(r)$, the inter-image liquid density profile along $\textbf{u}$ with $r$ the minimum distance from $\textbf{r}_{\tilde{i}}$ or $\textbf{r}_{\tilde{j}}^{\tilde{\textbf{p}}}$. In order to compute $\rho_{\textbf{u}}(r)$, we enumerate the average number of molecules that reside within cylindrical bins of radius $r_c$ and thickness 0.02~nm along $\textbf{u}$. We use a bin radius of $r_c$ not only to obtain better statistics, but also to assure that all the water molecules that are less than a hydration distance away from $\textbf{u}$ are included in the average. In order to obtain an accurate measure of density when either $\textbf{r}_{\tilde{i}}$ or $\textbf{r}_{\tilde{j}}^{\tilde{\textbf{p}}}$ are at the immediate vicinity of INP, each cylindrical bin is intersected with a plane parallel to the INP, and located at the smallest $z$ at which the supercooled liquid density becomes nonzero.   
}

\section{Results and Discussions}
\label{section:results}

\subsection{Summary of Nucleation Rates}
\label{section:summary}

\noindent
We first explore the dependence of nucleation rate on $L$, the dimension of the square-shaped INP. The computed rates are shown in Fig.~\ref{fig:schematics-rates}{\color{black}B} and listed in Table~\ref{table:rates}. We can, in particular, identify three distinct regimes for the dependence of rate on $L$. For small INPs,~i.e.,~for $L\le4.46$~nm, the rate is a strong function of $L$, and changes by as much as {\color{black}four} orders of magnitude, indicative of strong finite size effects and potentially spurious behavior. For large INPs,~i.e.,~for $L\ge7.65$~nm, the rate is a weak and monotonic function of $L$ and  increases by less than two orders of magnitude upon increasing $L$. These two regimes are highlighted in Figs.~\ref{fig:schematics-rates}{\color{black}B,C} with shaded green and red, respectively. There is, however, a third intermediate regime that is highlighted with a gradient shade in Figs.~\ref{fig:schematics-rates}{\color{black}B,C}, and that also exhibits a weak and monotonic dependence of rate on $L$. Unlike the second regime, however, the rate decreases slightly upon increasing $L$ in this intermediate regime.  

\begin{figure}
	\centering
	\includegraphics[width=.46\textwidth]{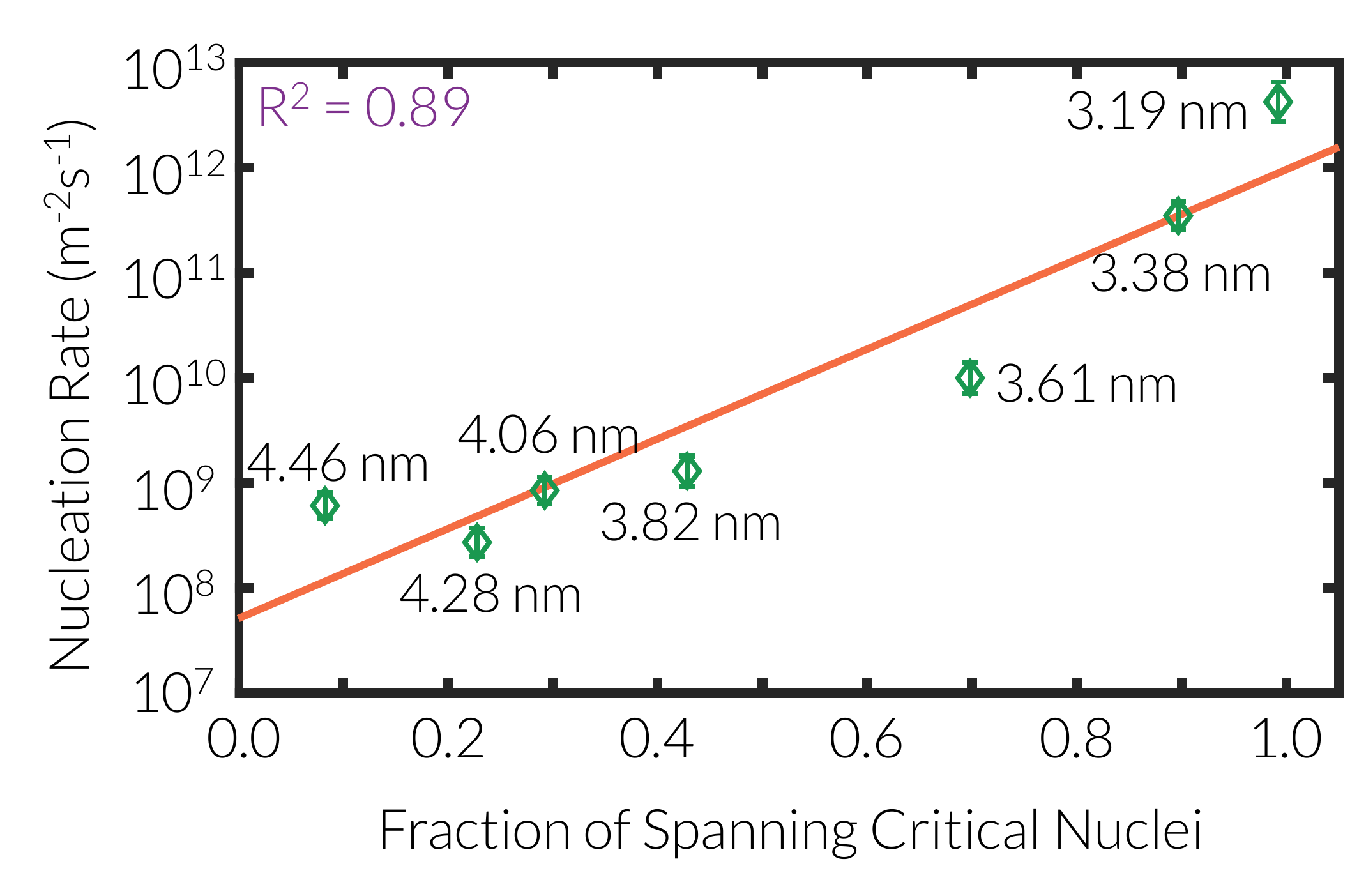}
	\caption{\label{fig:spanning-vs-rate}{\color{black}Correlation between the fraction of critical nuclei that span the simulation box and $\log_{10}\mathcal{J}$ in the spanning regime.}}
\end{figure}

In order to understand the origin of these contrasting behaviors, we focus on the geometric features of the critical nuclei. (See Section~\ref{section:jFFS} for our definition of critical configurations.) We first determine the fraction of the critical nuclei that {\color{black}are spanning using the approach described in Section~\ref{section:methods:spanning}.} Such spanning nuclei exhibit artificial directionality and spurious long range crystalline order along the $x$ and/or $y$ dimensions of the simulation box. Fig.~\ref{fig:schematics-rates}{\color{black}C} depicts the fraction of critical configuration with spanning crystalline nuclei. While such nuclei are prevalent for very small $L$'s, their fraction decreases upon increasing $L$ and eventually drops to zero at $L=5.74$~nm. {\color{black}Fig.~\ref{Fig:critical} depicts typical critical nuclei for systems of different sizes, including a spanning nucleus for $L=3.19$~nm and a non-spanning nucleus at $L=3.82$~nm.} The presence of an appreciable number of spanning critical nuclei is the main feature that distinguishes the small-INP regime from the other two. We will therefore refer to the shaded green region of Figs.~\ref{fig:schematics-rates}{\color{black}B,C} as the \emph{spanning regime}. The other two regimes, however, lack an appreciable fraction of spanning critical nuclei, and thus exhibit a weaker dependence of rate on $L$.

\begin{figure*}
\centering
\includegraphics[width=0.73\textwidth]{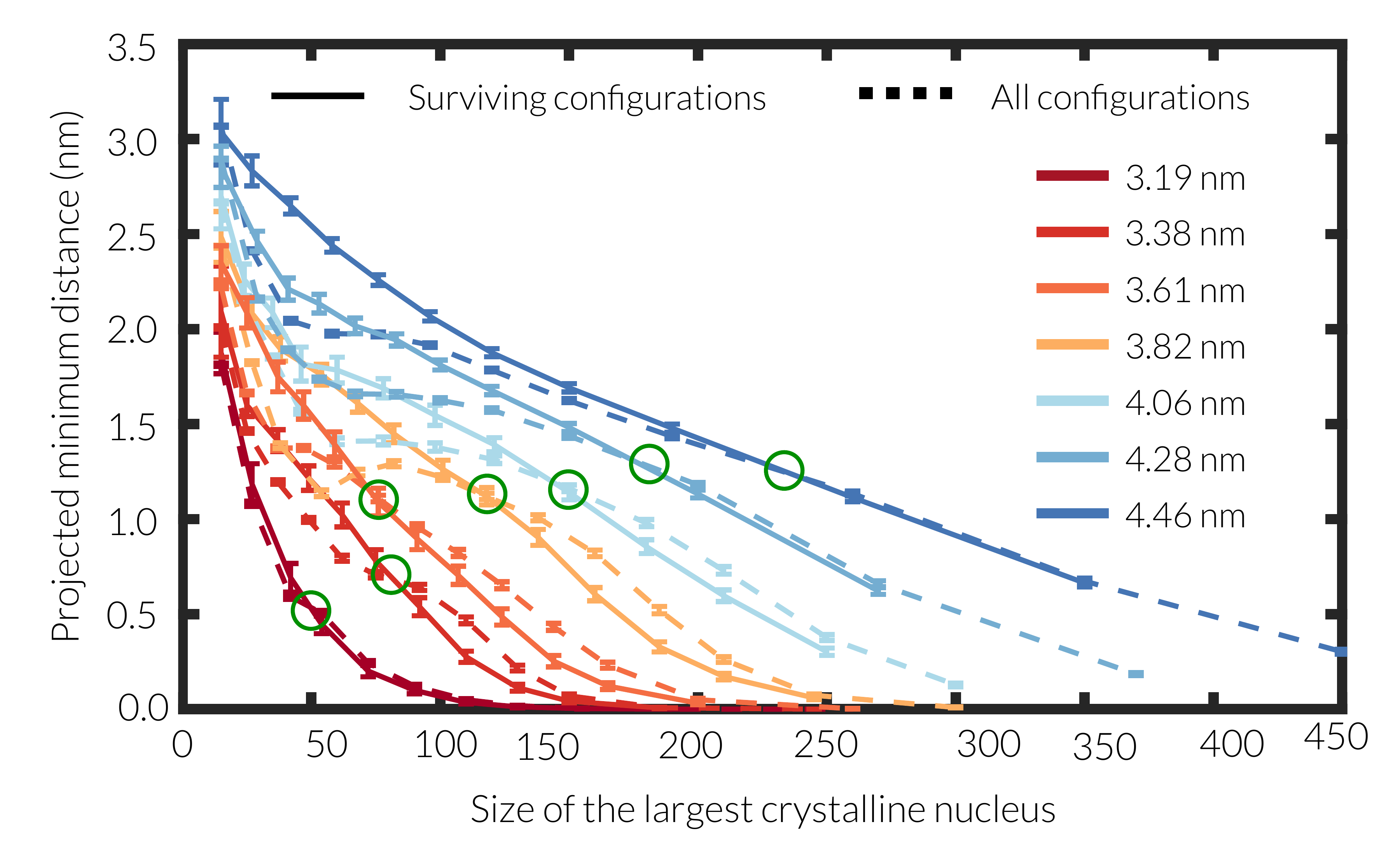} 
\caption{Average minimum projected  distances between the largest crystalline nuclei and their closest periodic images as a function of $\lambda$ for the configurations in the spanning regime. {\color{black}Green circles correspond to bottlenecks of the projected minimum distance.} Error bars correspond to 95\% confidence intervals.}
\label{fig:min-proj-dist}
\end{figure*} 

Finite size effects can, however, exist in the absence of spanning critical nuclei, as the proximity of a critical nucleus to its periodic image can still render noticeable finite size effects even in the absence of spanning. In order to devise a more rigorous measure of such proximity, {\color{black}we inspect the structure of the supercooled liquid that resides within inter-image regions of non-spanning nuclei (defined in Section~\ref{section:methods:spanning}).}  Fig.~\ref{fig:interimage-density} depicts {\color{black}the inter-image liquid density,} $\rho_{\textbf{u}}(r)$ for several system sizes where the spanning fraction is negligible. The inter-image liquid is fairly structured in the immediate vicinity of the nucleus, as evident in the three peaks of $\rho_{\textbf{u}}(r)$ for $r<0.77$~nm. {\color{black}The fact that a crystalline surface structures the liquid at its vicinity is a universal feature of crystal-liquid interfaces and is not limited to ice nucleation.} Note that  the heights and loci of these peaks are independent of the system size, so their emergence is unlikely to be impacted by finite size effects.  Beyond this threshold, which we denote by $r_{c,p}{\color{black}=0.77~\text{nm}}$, the liquid density reaches a plateau \textcolor{black}{vs.~$r$}. We therefore call the nuclei with $|\textbf{u}|\le 2r_{c,p}$ "proximal`` since the inter-image liquid confined between them and their closest periodic images clearly lacks bulk-like behavior. Since $|\textbf{u}|=0$ for a spanning cluster, all spanning clusters are also proximal.  Fig.~\ref{fig:schematics-rates}{\color{black}C} depicts the fraction of proximal critical nuclei as a function of $L$. Clearly, the presence of an appreciable fraction of proximal critical nuclei in the intermediate regime, and their absence in the large-INP regime is the main feature that distinguishes those two regimes. We  refer to the large-INP regime as the \emph{non-spanning regime}. The intermediate regime, however, constitutes a transition from lack of spanning and proximity in the non-spanning regime, to a preponderance of spanning and proximal critical nuclei in the spanning regime. We will analyze these three regimes separately and will explain in detail how the differences in the spanning behavior and proximity can affect the magnitude and the nature of finite size effects.

\subsection{The Spanning Regime}
\label{section:spanning}

\noindent
Due to the close proximity of critical nuclei and their periodic images in the spanning regime, the kinetics and mechanism of nucleation exhibit a strong dependence on system size.  Overall, we observe that an increase in the fraction of spanning configurations results in an increase in rate (Fig.~\ref{fig:schematics-rates}{\color{black}B}) and a decrease in the size of the critical nucleus (Table.~\ref{table:rates}). This indicates that the ability to form spanning nuclei artificially promotes nucleation. {\color{black}There is indeed a reasonably good-- but imperfect-- correlation between the fraction of spanning configurations and $\log_{10}\mathcal{J}$ as can be seen in Fig.~\ref{fig:spanning-vs-rate}. This imperfect correlation, however, suggests that  collective variables other than the spanning fraction might be better predictors of the nucleation rate in the spanning regime. At a fundamental level, such a predictor needs to account for the effect of proximity between crystalline nuclei and their periodic images prior to the occurrence of spanning.  One such suitable candidate is}   $d_{\text{min}}^\text{proj}=|(u_x,u_y,0)|_2$, or the minimum $xy$-projected distance between a crystalline nucleus and its closest periodic image.  {\color{black}Note that $d_{\text{min}}^{\text{proj}}=0$ for a spanning nucleus.} Focusing on this lateral distance enables us to detect and quantify peculiarities that arise {\color{black}in the absence of} spanning.  Fig.~\ref{fig:min-proj-dist} shows $ d_{\text{min}}^\text{proj}$ as a function of crystalline nucleus size for all seven films in the spanning regime. In addition to all configurations at a given milestone, we compute $d_{\text{min}}^\text{proj}$ for the surviving configurations as well.  As expected, both $d_{\text{min},s}^{\text{proj}}(\lambda)$ and $d_{\text{min},a}^{\text{proj}}(\lambda)$ are strictly decreasing functions of $\lambda$, but their decline is faster in smaller films. At earlier FFS milestones and prior to the bottlenecks (the green circles in Fig.~\ref{fig:min-proj-dist}), however, $d_{\text{min},s}^{\text{proj}}(\lambda)$ is consistently larger than $d_{\text{min},a}^{\text{proj}}(\lambda)$ irrespective of the system size.  This suggests that at  initial stages of nucleation, a less spread-out nucleus is better suited to survive and to contribute to the nucleation pathway. Beyond the bottlenecks, which are always smaller than $N^*$, $d_{\text{min},s}^{\text{proj}}(\lambda)$ remains consistently smaller than $d_{\text{min},a}^{\text{proj}}(\lambda)$. This suggests that the more compact surviving nuclei of earlier milestones reach a geometry at the bottleneck that facilitates their further growth towards their periodic images to form spanning or proximal critical nuclei.  The bottlenecks of $d_{\min}^{\text{proj}}(\cdot)$ therefore constitute {\color{black}pivotal stages} in the nucleation process, and the properties of the crystalline nuclei therein are likely to play an important role in the kinetics and mechanism of nucleation.

\begin{figure*}
	\centering
	\includegraphics[width=.7\textwidth]{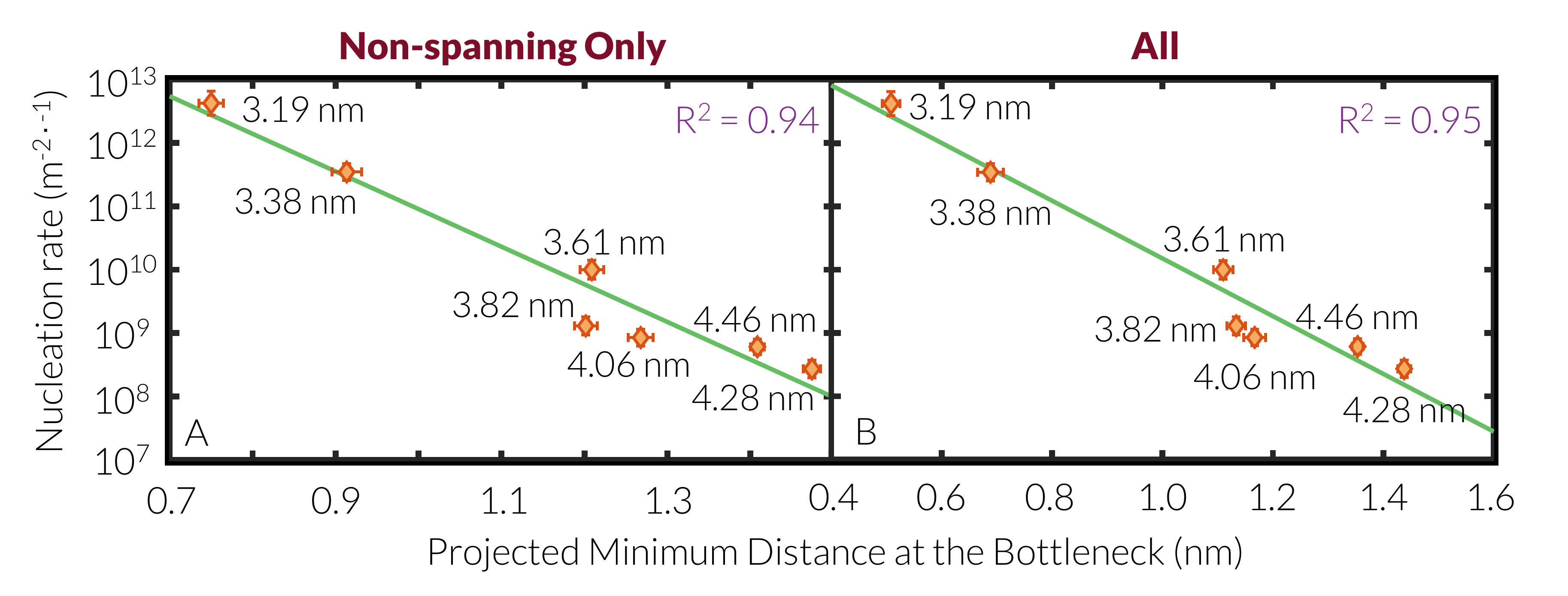}
	\caption{\label{fig:proj-min-dist}
	{\color{black} Correlation between the minimum projected  distance of (A) non-spanning and (B) all configurations at the bottleneck vs.~$\log_{10}\mathcal{J}$ in the spanning regime.
	}
	}
\end{figure*}

\begin{figure}
\centering
\includegraphics[width=0.44\textwidth]{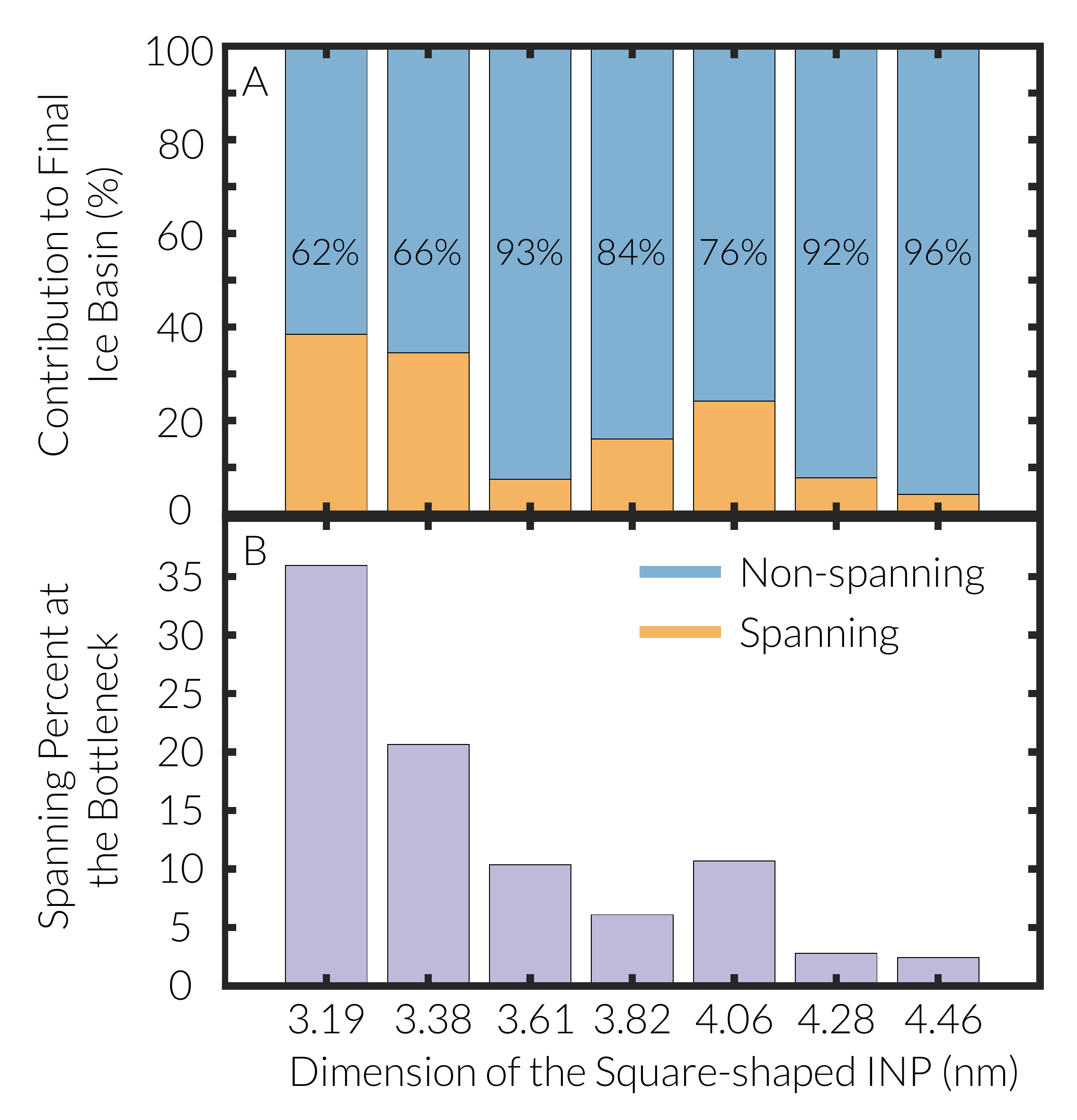} 
\caption{(A) Percentage of configurations at the last milestone that originate from the spanning and non-spanning configurations at the bottleneck. (B) Percentage of configurations at the bottleneck with spanning crystalline nuclei.}
\label{fig:spanning-bottleneck}
\end{figure}

{\color{black} This hypothesis is borne out by the strong correlation between $d_{\text{min},b}^{\text{proj}}$ and $\log_{10}\mathcal{J}$ as depicted in Fig.~\ref{fig:proj-min-dist}. The correlation is equally strong whether $d_{\text{min},b}^{\text{proj}}$ is computed only for the non-spanning configurations (Fig.~\ref{fig:proj-min-dist}A) or all configurations (Fig.~\ref{fig:proj-min-dist}B) at the bottleneck. 
This is not surprising since the overwhelming majority of crystalline nuclei at the bottleneck are non-spanning as can be seen in Fig.~\ref{fig:spanning-bottleneck}B. Moreover, a large fraction of configurations at the final crystalline basin emanate from non-spanning surviving configurations at the bottleneck as can be seen in Fig.~\ref{fig:spanning-bottleneck}A. It must, however, be noted that while spanning is much less pronounced at the bottlenecks, it can still play a role in determining the kinetics of nucleation. Indeed,  including the spanning configurations in the average, $d_{\text{min},b}^{\text{proj}}$,  slightly improves the fit as can be seen in Fig.~\ref{fig:proj-min-dist}B. 
} 

\begin{figure*}
\centering
\includegraphics[width=0.72\textwidth]{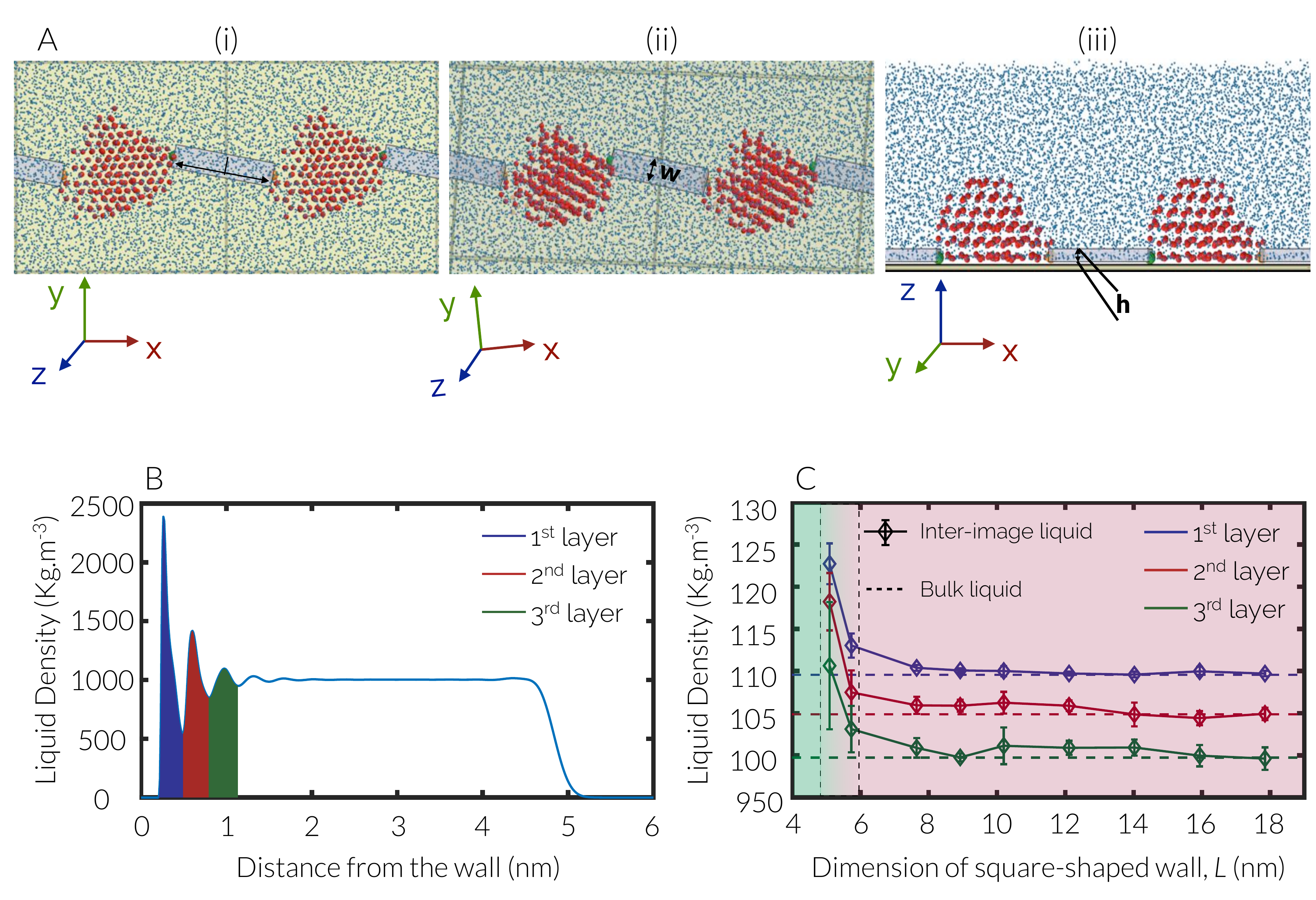} 
\caption{(A) Three different views of a typical cuboid along $\textbf{u}_{xy}=(u_x,u_y,0)$. The water molecules connected by $\textbf{u}$ are both in the first layer and are represented with orange and green spheres, respectively. (B) The density profile of the supercooled liquid in the vicinity of the largest INP at 235~K. The first three peaks of density are highlighted with different colors. (C) Average densities of the inter-image liquid in the plateau region of Fig.~\ref{fig:interimage-density}, computed only for the configurations in which both $\tilde{i}$ and $\tilde{j}$-- orange and green spheres in (A)-- are located in the same layer. }
\label{fig:nonspanning-density}
\vspace{-10pt}
\end{figure*}

\subsection{Non-spanning and Intermediate Regimes}\label{section:non-spanning}

\noindent
As mentioned in Section~\ref{section:summary}, the non-spanning regime is characterized by critical nuclei that are neither spanning nor proximal, while the intermediate regime contains an appreciable fraction of proximal but non-spanning critical nuclei. More precisely, 19\% and 1.8\% of the critical nuclei within the two systems that belong to the intermediate regime, namely $L=5.10$ and $5.74$ nm, are proximal, respectively. (A tiny fraction of critical nuclei (0.2\%) in the smaller 5.10-nm system are spanning.) It can thus be argued that the dependence of rate on $L$ in these two regimes is likely governed by the properties of the supercooled liquid within the plateau region of Fig.~\ref{fig:interimage-density} {\color{black} considering the prevalence of non-proximal configurations in both regimes}.

In order to test this hypothesis, we compute the average inter-image liquid density within the plateau region of $\rho_{\textbf{u}}(r)$ and compare it to the density of the supercooled liquid under the same conditions. {\color{black}Like other solid surfaces, however, the INP considered in this work structures the liquid at its vicinity as can be seen in Fig.~\ref{fig:nonspanning-density}B. In particular, there are three distinct liquid layers near the INP, with average densities different from that of the bulk liquid.} As such,  the average liquid density within the inter-image plateau region should  be compared to the average liquid density along $\textbf{u}$, which can, in general, be different from the bulk density. {\color{black}This comparison is, however, not trivial when the inter-image vector $\textbf{u}$ passes through multiple liquid layers.} We therefore simplify our analysis by computing inter-image plateau densities for configurations {\color{black}in which $\textbf{u}$ is fully located within a single liquid layer.} Indeed, around 50\% of all critical configurations in the non-spanning and intermediate regimes have {\color{black}inter-image vectors} lying within {\color{black}one of} the first three liquid layers. For {\color{black}any such configuration,}
the inter-image plateau density is computed by enumerating $N_p$, the number of water molecules that reside within a cuboid of length $d_{\min}^{\text{proj}}-2r_{c,p}$, width $w=0.64$~nm and height $h$ along $\textbf{u}_{xy}=(u_x,u_y,0)$ and centered at the midpoint between the {\color{black}nucleus} and its closest periodic image. We choose $w$  so that all in-layer nearest neighbors of molecules along $\textbf{u}_{xy}$ are included within the cuboid, while $h$ is chosen as the thickness of {\color{black}the liquid layer,} $\mathcal{L}$ {\color{black}determined from} Fig.~\ref{fig:nonspanning-density}B. A typical cuboid for a configuration with a $\textbf{u}$ within the first liquid layer is shown from three different angles in Fig.~\ref{fig:nonspanning-density}A. The average inter-image plateau density for each layer is then estimated as the weighted average of individual densities $\rho_p(x)=N_p(x)/wh[d_{\min}^{\text{proj}}(x)-2r_{c,p}]$, and is given by,
\begin{eqnarray}
\rho_p^{\mathcal{L}} &=& \frac{\sum_{x\in\mathcal{L}}N_p(x)\rho_p(x)}{\sum_{x\in\mathcal{L}}N_p(x)},
\end{eqnarray} 
{\color{black}where $x\in\mathcal{L}$ is a configuration whose $\textbf{u}$ fully lies within $\mathcal{L}$.} {\color{black}Note that weighting enhances the statistical accuracy of $\rho_p^{\mathcal{L}}$ by favoring configurations with larger inter-image  plateau regions.}
The average liquid density within $\mathcal{L}$ is estimated {\color{black}by integrating the density profile of Fig.~\ref{fig:nonspanning-density}B within each layer.} 

Fig.~\ref{fig:nonspanning-density}C depicts the $\rho_p^{\mathcal{L}}$'s of the three layers highlighted in Fig.~\ref{fig:nonspanning-density}B for the critical configurations in the intermediate and non-spanning regimes. For the two films in the intermediate regime, the inter-image plateau densities are significantly larger than the per-layer supercooled liquid densities. This effect likely arises from the negatively sloped melting curve of water,~i.e.,~the fact that the liquid is denser than the crystal under ambient conditions. More precisely, the incorporation of new water molecules into the growing crystalline nucleus results in an increase in its volume, and the compression of the inter-image liquid during the out-of-equilibrium nucleation process. Since the thermodynamic driving force for crystallization is a decreasing function of density in water, larger inter-image densities will result is larger nucleation barriers and smaller nucleation rates.  While this effect is partly offset by the existence of an appreciable fraction of proximal critical nuclei, it results in rates that are generally smaller than those in the other two regimes. {\color{black}It is necessary to emphasize that this effect is not an artifact of using the canonical ensemble as the free interface is able to to freely adjust to density changes during nucleation.}

This difference almost disappears in the non-spanning regime, presumably due to the larger size of the inter- image region, which better "absorbs`` the growing nucleus front. Indeed, the inter-image plateau densities are statistically indistinguishable from the corresponding liquid densities for $L\ge{\color{black}7.65}$~nm. These findings generally indicate that the inter-image region is large enough to not be impacted by strong finite size effects in the non-spanning regime. While these findings are based on the configurations with {\color{black}inter-image vectors} entirely within one of the first three layers, we expect a qualitatively similar behavior if configurations with inter-layer minimum image connections are also included.

\begin{figure}
\centering
\includegraphics[width=.49\textwidth]{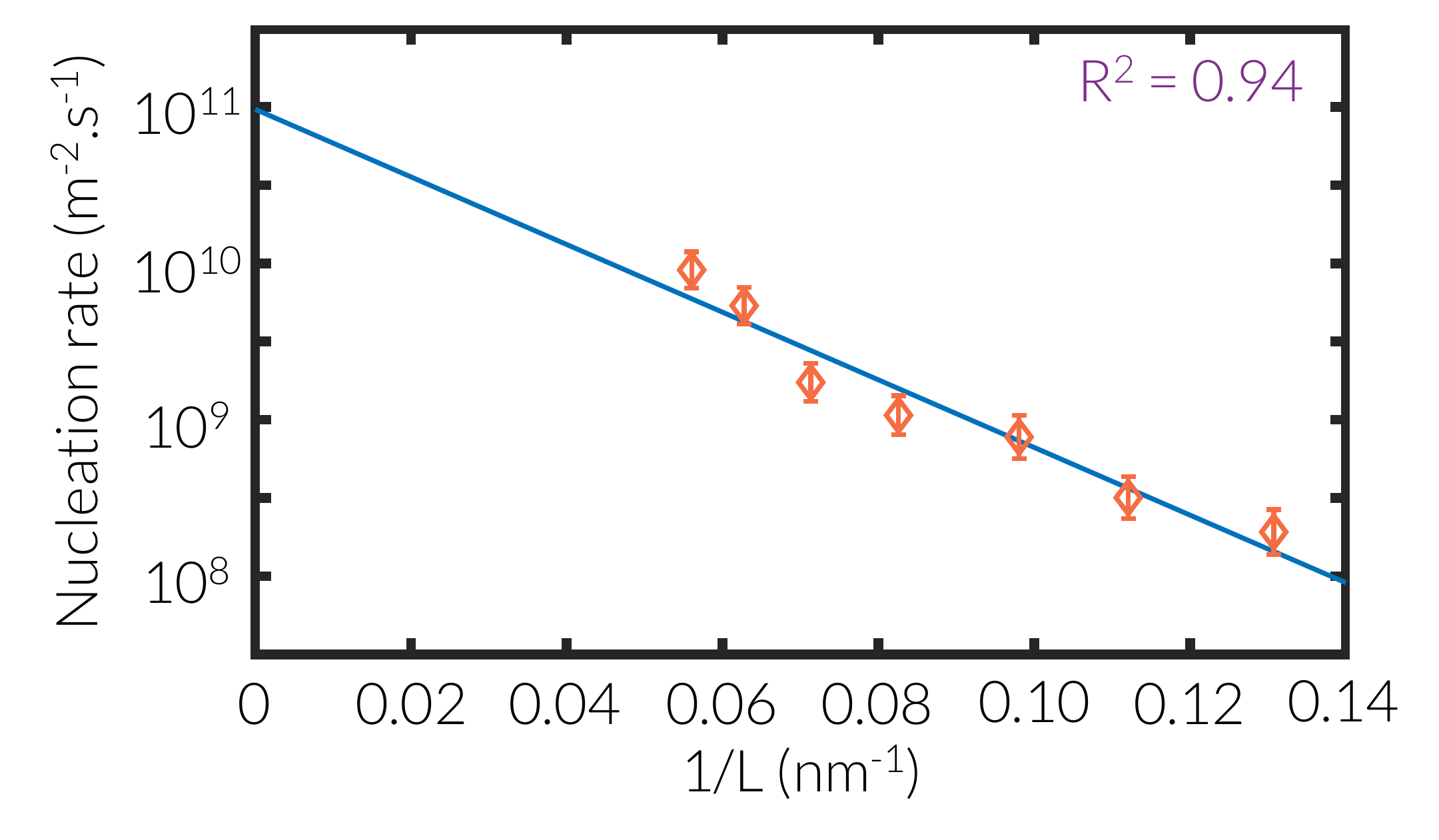}
\caption{Linear correlation between $\log_{10}\mathcal{J}$ and $1/L$, indicative of a weak finite size effects in the non-spanning regime.}
\label{fig:scaling}
\end{figure}

{\color{black}\subsubsection{Linear Scaling of $\log_{10}\mathcal{J}$ with $1/L$ in the Non-spanning Regime}}

\noindent
Despite lack of any quantifiable structural signatures, the nucleation rate is still a weak function of $L$ in the non- spanning regime. Indeed, the rate increases by around 1.{\color{black}7} orders of magnitude upon increasing $L$ from 7.65~nm and 17.85~nm. The dependence of rate on system size can be satisfactorily described using a linear correlation between $\log_{10}\mathcal{J}$ and $1/L$, as depicted in Fig.~\ref{fig:scaling}. 
It has indeed been reported that the averages of properties such as  critical temperature,\cite{MonJChemPhys1992} free energy difference,\cite{MastnyJChemPhys2007} structure factor,\cite{SalacusePhysRevE1996} diffusivity,\cite{JamaliJChemTheoryComput2018} surface tension\cite{AguadoJChemPhys2001} and percolation threshold\cite{HajiAkbariPRE2009} have a power law dependence on the system size. The linear relationship between $\log_{10}\mathcal{J}$ and $1/L$ can therefore be rationalized by invoking the formalism of classical nucleation theory (CNT)~\cite{TurnballJCP1950} according to which the nucleation rate is given by,
\begin{eqnarray}
\mathcal{J} &=& A\exp\left[-\frac{\Delta{G}_{\text{nuc}}}{k_BT}\right],
\end{eqnarray}
Here, $A$, $k_B$ and $T$ are the kinetic prefactor, the Boltzmann constant, and temperature, respectively. $\Delta{G}_{\text{nuc}}$ is the nucleation barrier and is given by:
\begin{eqnarray}\label{eq:Delta-G-CNT}
\Delta{G}_{\text{nuc}} &=& \frac{4\pi\gamma_{sl}^3(1-\cos\theta)^2(2+\cos\theta)}{3\rho_s^2|\Delta\mu|^2},
\end{eqnarray}
with $\gamma_{sl}, \Delta\mu, \rho_s$ and $\theta$, the solid-liquid surface tension, the chemical potential difference between the two phases, the solid number density, and the solid-liquid-INP contact angle, respectively. All these quantities are known to be impacted by finite size effects,\cite{AguadoJChemPhys2001, OreaJCP2005, MastnyJChemPhys2007,  BiscayJCP2009, BurtJPhysChemC2016} and some of them exhibit a power-law dependence on system size.\cite{AguadoJChemPhys2001, MastnyJChemPhys2007} Consequently, the linear correlation between $\log_{10}\mathcal{J}$ and $1/L$ is a likely consequence of the power-law dependence of some of these underlying thermodynamic properties {\color{black}as demonstrated in Appendix~\ref{appendix:section:power-law}}. This scaling enables us to use the rates computed in sufficiently large finite systems to estimate the heterogeneous nucleation rate in the thermodynamic limit. According to the correlation depicted in Fig.~\ref{fig:scaling}, the nucleation rate at the thermodynamic limit ($L\rightarrow\infty$) is given by $\log_{10}\mathcal{J}_{\infty}={\color{black}10.99\pm0.58}$. 

\vspace{15pt}

{\color{black}

\section{A Proposed Heuristic to Assess and Quantify Finite Size Effects in Crystal Nucleation Studies}
\label{section:heuristics}

\noindent 
We use our observations to propose a set of heuristics for assessing whether a particular {\color{black}nucleation} rate calculation is impacted by finite size artifacts. These heuristics constitute a decision tree comprised of the following steps:
\begin{enumerate}
	
	\item Determine $N^*$, the critical nucleus size, using an approach such as the mean first passage time (MFPT) method\cite{WedekindJChemPhys2007} or committor analysis.\cite{PetersJCP2006} Note that the particular classification and clustering algorithm utilized for detecting crystalline nuclei will impact the precise  value of $N^*$ but will not affect whether a particular configuration is classified as critical,~i.e.,~whether it has a commitor probability close to 50\%.
	
	\item Obtain $\mathcal{C}^*$, a collection of critical configurations, by analyzing long MD trajectories, or scanning the configurations collected at FFS or transition interface sampling (TIS)\cite{VanErpJCompPhys2005} milestones. Ideally, the largest crystalline nucleus in each such configuration should be comprised of exactly $N^*$ molecules. In practice though, such a strict filter will limit the number of configurations in $\mathcal{C}^*$, and will, in turn, lead to poor statistics. Therefore, a less strict criterion can be adopted for constructing $\mathcal{C}^*$,~e.g.,~that the committor probability of each configuration is between 30\% and 70\%. 
	
	\item Use the approach described in Section~\ref{section:methods:spanning} to determine whether any configuration in $\mathcal{C}^*$ spans across the periodic boundary, in which case the rate calculation is likely impacted by strong finite size effects. Note that the utilized classification and clustering algorithm can significantly impact the fraction of spanning configurations, or whether any spanning configuration is detected at all. Therefore, while the presence of spanning configurations is a strong indication of finite size artifacts, their absence does not rule out the existence of such artifacts. 	
	
	\item Determine the inter-image vector $\textbf{u}$ for every non-spanning configuration in $\mathcal{C}^*$ and compute the profile of one (or several) relevant structural properties along $\textbf{u}$ by averaging over all non-spanning configurations. In most systems,  density is the most relevant quantity to monitor. Depending on the nature of the liquid and crystalline phases, however, other structural features might be more relevant. Examples include charge density in systems of polyelectrolytes, ionic liquids or solutions, local orientational order parameters\cite{HajiAkbariJPhysA2015} in systems of rigid or semi-rigid anisotropic building blocks, or composition in mixtures and solutions. If any of these profiles fails to converge to a plateau, the rate calculation is likely impacted by strong finite size effects. Otherwise, identify the decay length scale for each profile, and denote the nuclei with $|\textbf{u}|$'s shorter than twice the largest decay length scale as proximal.  If an appreciable fraction of configurations in $\mathcal{C}^*$ are proximal, the calculation is likely affected by finite size artifacts.  It is necessary to emphasize that whether a particular configuration is proximal or not is not very sensitive to the classification and clustering algorithm employed for detecting crystalline nuclei. Therefore, proximity of a nucleus to its closest periodic image is a more robust measure of finite size artifacts than its apparent spanning-- or lack thereof-- the periodic boundary. 		
	
	\item If only a tiny fraction of configurations in $\mathcal{C}^*$ are proximal, compute the average inter-image  plateau value for each profiled property, and compare it with that of the liquid under the same conditions. If no statistically significant difference is observed, the calculation is likely free of strong finite size effects. As discussed in Section~\ref{section:non-spanning}, all liquid properties (including density) become functions of position in the presence of an interface. Therefore in the case of heterogeneous nucleation, not only liquid properties will be functions of position, but also the inter-image plateau averages will depend on the location of the inter-image vector, and any comparison between the two should take this into consideration.
	
\end{enumerate}

Considering the limited scope of the calculations conducted in this work (heterogeneous ice nucleation on a structureless INP using a coarse-grained model of water), it is not clear whether the heuristics proposed here can be universally applied to crystal nucleation (homogeneous and heterogeneous) in other systems. Even in the case of ice nucleation, finite size effects might play out differently if an atomistic model (such as TIP4P/Ice) is utilized, or if the underlying INP is structured.  Therefore, further follow-up studies are necessary to assess the robustness of these heuristics. Even if these heuristics are proven to be robust, there are two major outstanding challenges still to be addressed. First of all, it is not always \emph{a priori} clear what structural features are relevant for quantifying proximity. While potential candidates can be surmised by considering the symmetries and composition of the underlying liquid, it is always possible that  hidden structural features with long correlation lengths  impact nucleation even in very large systems. The second challenge is that the extent by which the rate scales with system size might also vary considerably depending on the system type (e.g.,~the presence of long-range electrostatic interactions, or molecules with long-range connectivity such as polymers) and the magnitude of the thermodynamic driving force. This can be particularly problematic in the non-spanning regime where the proposed heuristic will predict minimal finite size effects, but the rate might still change considerably upon changing the system size.

We wish to note that even though these heuristics are to be applied to the findings of a \emph{completed} rate calculation, they can be utilized in conjunction with other methods such as seeding\cite{SanzJACS2013} to decide the proper size of a simulation box \emph{before} a rate calculation is conducted. More precisely, one can create seeds of different sizes of a target crystal within a bath of supercooled liquid (or supersaturated solution), and equilibrate them using a method such as umbrella sampling\cite{TorrieJCompPhys1977} or hybrid Monte Carlo.\cite{DuanePhysLettB1987, GuoJTheorComputChem2018} The structure of the  liquid within the inter-image regions can then be inspected to determine the proper decay length scales for relevant liquid properties (such as density). The approximate size of a critical nucleus can then be determined separately from the seeding method. Assuming that the characteristic length scale of a critical nucleus, and the largest decay length scale of all profiled liquid properties are given by $l_{\text{cr}}$ and $l_{d}$, respectively, a simulation box with dimensions sufficiently larger than $l_{\text{cr}}+2l_d$ will likely be free of strong finite size effects. }

\section{Conclusion}
\label{section:conclusion}

\noindent
In this work, we use MD simulations and jumpy forward flux sampling to explore the sensitivity of the heterogeneous ice nucleation kinetics and mechanism to $L$, the size of a model structureless INP, and find that finite size effects that arise due to periodic boundary conditions can significantly affect heterogeneous nucleation rates over a wide range of INP sizes. We observe that {\color{black}in this particular system,} nucleation rates  change by as much as {\color{black}four} orders of magnitude within the range of $L$'s considered in this work, and are also non-monotonic functions of $L$. We identify three distinct regimes for the dependence of rate on $L$ based on whether critical crystalline nuclei span across the periodic boundary, and if not, whether they are proximal, i.e.,~that the liquid that separates them from their closest periodic images is fully structured. In the spanning regime, an appreciable fraction of critical nuclei span across the periodic boundary, while the overwhelming majority of those that do not span are proximal. This results in a strong dependence of rate  on $L$, which we explain by analyzing the average minimum projected distance of subcritical crystalline nuclei at the bottleneck FFS milestones. We demonstrate that the fraction of spanning bottleneck configurations {\color{black}and} the average minimum projected distance for the non-spanning ones collectively explain the variations in rate in the spanning regime. The second regime, which is observed for intermediate-sized INPs and is thus called the intermediate regime, is characterized by the emergence of proximal-- but non-spanning-- critical nuclei. Within this regime, the rate is a weak function of $L$, and its variations is governed by the fraction of critical nuclei that are proximal, and the inter-image plateau liquid density value for those that are not. In the third regime, which is observed for large INPs and which we denote as the non-spanning regime, critical nuclei are neither spanning nor proximal, and the rate is a weakly increasing function of $L$. We are able to demonstrate that there is a linear scaling between $\log_{10}\mathcal{J}$ and $1/L$, which we use to estimate the nucleation rate in the thermodynamic limit. 

The key heuristic that emerge from this work is that finite size effects are minimal or absent in systems where the critical nuclei are neither spanning nor proximal, and the inter-image liquid {\color{black}profiles} in the plateau region do not deviate significantly from that of the liquid under the same conditions. {\color{black}A more algorithmic description of these heuristics is outlined in Section~\ref{section:heuristics}. Note that} these  criteria  can be unambiguously tested for both homogeneous and heterogeneous crystal nucleation in all systems and irrespective of the particular definition of the order parameter. While whether a nucleus spans across the periodic boundary will depend on the classification algorithm utilized for detecting solid-like molecules, its proximity to its closest periodic image-- or lack thereof-- and its inter-image plateau density-- if it is not proximal-- are independent of such details. It must be noted that the particular system size beyond which these conditions are satisfied will depend on the type of the system and the thermodynamic state. It might therefore be risky to use \emph{ad hoc} heuristics, such as the "10\% rule`` discussed in Section~\ref{section:intro} to determine whether a rate calculation is impacted by finite size effects. Determining how these heuristics translate into proper system sizes in different systems can be the topic of future studies.  Moreover, even when these conditions are satisfied, the rate can still be a weak function of the system size, presumably due to the size dependence of the underlying thermodynamic properties. Such a dependence might be particularly strong in systems with long-range electrostatic interactions.

{\color{black}
As discussed in Section~\ref{section:heuristics}, the heuristics proposed in this work are expected to be applicable to homogeneous nucleation studies as well. The fact that homogeneous nucleation occurs endogenously within a liquid makes it simpler to analyze, as no particular shape or polymorph composition will be exerted onto the crystalline nuclei by an external interface. Moreover, due to the translational isotropy of bulk liquids, averages of structural properties such as density will not be position-dependent and can thus be readily calculated with high statistical precision. This makes comparing inter-image plateau properties with those of the bulk liquid more straightforward. On the other hand, validating these heuristics for homogeneous nucleation will be more expensive computationally not only because  homogeneous nucleation rate are generally much smaller, but also because the system size scales as $L^3$ (rather than $L^2$) with $L$ the characteristic dimension of the simulation box.
}

{\color{black}

It is widely known that nucleation rate is extremely sensitive to the mathematical form and the parameters of the employed force-field, as well as thermodynamic variables, such as pressure, temperature and concentration. Even subtle changes in any of these parameters can alter the nucleation rate considerably, sometimes by as much as tens of orders of magnitude.\cite{GianettiPCCP2016}  It might therefore be argued that finite size effects are not that significant in comparison, as the ensuing errors appear to be modest. It is not, however, clear that errors due to finite size effects will always be limited to a few orders of magnitude (as observed in this work). Instead, it is totally plausible that the extent by which nucleation rates in finite systems deviate from those in the thermodynamic limit might depend heavily on the magnitude of the nucleation barrier. It is therefore important to determine the conditions under which finite size effects will be minimal. Otherwise, it will be difficult to trust the accuracy of rates computed as a function of quantities such as temperature, electric field, concentration and force-field parameters, particularly when the computed rates differ by many orders of magnitude and exhibit anomalous or nonclassical behavior. In addition, there are  instances in which one might need to compute the nucleation rate in systems with different geometries but at a single thermodynamic state point and using the same force-field. 
Since the underlying thermodynamic properties remain mostly unchanged under such circumstances, the ensuing changes in rate are usually modest, and resolving them accurately requires having robust tools to assess whether the corresponding calculations are impacted by finite size effects.}

The strong sensitivity of rate to $L$ in the spanning regime has important implications for detecting finite size artifacts in computational studies of nucleation, as it points to the unpredictable nature of such effects when the system size is too small. Indeed, the rates computed for certain $L$'s in this regime are fortuitously identical to those computed for much larger $L$'s in the other two regimes. This suggest that the common practice of validating nucleation rates by comparing them to rates computed for a larger system might yield misleading results since both systems might be affected by finite size effects-- but for different reasons. 

\section*{Supplementary Material}
\noindent
Technical details of the conducted FFS calculations, including the locations of milestones, and the number of trial trajectories and successful crossings for each iteration are included in the supplementary material.

\begin{acknowledgments}
\noindent A.H.-A. gratefully acknowledges the support of the National Science Foundation CAREER Award (Grant No. CBET-1751971). These calculations were performed on the Yale Center for Research Computing. This work used the Extreme Science and Engineering Discovery Environment (XSEDE), which is supported by National Science Foundation grant no. ACI-1548562~\cite{TownsCompSciEng2014}. 
\end{acknowledgments}

\section*{Data Availability Statement}

\noindent
The underlying data supporting the findings of this study are available from the corresponding author upon reasonable request.

\vspace{-5pt}

\appendix 

{\color{black}
\section{Power Law Scaling of $\log\mathcal{J}$ with $L$}
\label{appendix:section:power-law}

\noindent
Let the physical properties in Eq.~(\ref{eq:Delta-G-CNT}) have the following power-law dependence on $L$,
\begin{subequations}
\begin{eqnarray}
\gamma_{sl}(L) &=& \gamma_{sl,\infty}(1+aL^{-\alpha})\\
\theta(L) &=& \theta_\infty(1+bL^{-\beta})\\
\left|\Delta\mu(L)\right| &=& \left|\Delta\mu_\infty\right|(1+pL^{-\nu})\\
\rho_s(L) &=& \rho_{s,\infty}(1+kL^{-\kappa})
\end{eqnarray}
\end{subequations}
with $\alpha,\beta,\nu,\kappa>0$ and $a,b,p,k$ real-valued constants. Note that we consider the most general case in which all model parameters in CNT have a power-law dependence on $L$ even though such behavior is not established for all of them. We would, however, note that the analysis presented here can be easily revised if some of these exponents are zero, without changing our final conclusion. Using the Taylor expansion $(1+x)^q\sim 1+qx$ yields,
\begin{subequations}
\begin{eqnarray}
\frac{\gamma_{sl}^3(L)}{ \gamma_{sl,\infty}^3} &\approx&1+3aL^{-\alpha}\label{eq:gamma-inf}\\
\frac{\left(1-\cos\theta\right)^2(2+\cos\theta)}{\left(1-\cos\theta_\infty\right)^2(2+\cos\theta_\infty)} &\approx& 1 + b'L^{-\beta}\\
\left|\frac{\Delta\mu(L)}{\Delta\mu_\infty}\right|^2 &\approx& 1+2pL^{-\nu}\\
\frac{\rho_s^2(L)}{\rho_{s,\infty}^2} &\approx& 1+2kL^{-\kappa}
\end{eqnarray}
\end{subequations}
where $b'$ is given by,
\begin{eqnarray}
b '&=& \frac{3b\theta_\infty\sin^3\theta_\infty}{(1-\cos\theta_\infty)^2(2+\cos\theta_\infty)}
\end{eqnarray}
These scaling relationships can thus be used to determine the scaling of $\Delta{G}_{\text{nuc}}$ with $L$:
\begin{eqnarray}
\frac{\Delta{G}_{\text{nuc}}(L)}{\Delta{G}_{\text{nuc},\infty}} &\approx & \frac{(1+3aL^{-\alpha})(1+b'L^{-\beta})}{(1+2pL^{-\nu})(1+2kL^{-\kappa})} \notag\\
&\approx& (1+3aL^{-\alpha})(1+b'L^{-\beta}) \notag\\
&& \times (1-2pL^{-\nu})(1-2kL^{-\kappa}) \label{eq:DeltaGnuc-vs-L}
\end{eqnarray}
Note that the leading $L$-dependent term in Eq.~(\ref{eq:DeltaGnuc-vs-L}) scales as $L^{-\min\{\alpha,\beta,\nu,\kappa\}}$. Therefore, $\log_{10}\mathcal{J}$ will have a power-law dependence on $L$. The linear scaling observed in Fig.~\ref{fig:scaling} is a special case of this power-law behavior and will become possible if $\min\{\alpha,\beta,\nu,\kappa\}\approx1$. 
}

\bibliographystyle{apsrev}
\bibliography{References}

\appendix 

\section*{SUPPLEMENTARY INFORMATION}

\setcounter{figure}{0}
\setcounter{table}{0}
\renewcommand{\thefigure}{S\arabic{figure}}
\renewcommand{\thetable}{S\arabic{table}}

\begin{table*}
\caption{Technical specifications of basin exploration simulations. All calculations are conducted at 235~K using the mW model.}
\centering
\begin{tabular}{c c c r r r c c}
\hline\hline
$L$~[nm] & ~~$\lambda_A$~~ & ~~$\lambda_0$~~ & $N_0$~~~~ & ~~$\mathcal{T}$~[ns]~~~~ & $S=L^2$~[$\text{nm}^2$] & ~~$\Phi_0 [\text{m}^{-2}\cdot\text{s}^{-1}] $~~ & ~~$\log_{10}\Phi_0~[\text{m}^{-2}\cdot\text{s}^{-1}]$~~\\ [0.5ex]
\hline
3.1869 & 4 & 15 & 5,802~~ & 1,507.7~~ & 10.157~~~~ & $3.7888 \times 10^{26}$ & +26.5785 \\
3.3803 & 4 & 15 & 6,103~~ & 1,388.8~~ & 11.426~~~~ & $3.8460 \times 10^{26}$ & +26.5850 \\
3.6056 & 4 & 15 & 5,414~~ & 1,086.4~~ & 13.001~~~~ & $3.8331 \times 10^{26} $ & +26.5836 \\
3.8243 & 4 & 15 & 9,379~~ & 1,703.2~~ & 14.626~~~~ & $3.7650 \times 10^{26}$ & +26.5758\\
4.0563 & 4 & 15 & 6,584~~ & 1,068.2~~ & 16.454~~~~ & $3.7460 \times 10^{26}$ & +26.5736 \\
4.2817 & 4 & 15 & 18,178~~ & 2,624.5~~ & 18.333~~~~ & $3.7780 \times 10^{26}$ & +26.5773 \\
4.4617 & 4 & 15 & 18,741~~ & 2,554.5~~ & 19.907~~~~ & $3.6853 \times 10^{26}$ & +26.5665 \\
5.0991 & 4 & 15 & 9,802~~ & 1,010.2~~ & 26.001~~~~ & $3.7318 \times 10^{26}$  & +26.5719 \\
5.7365 & 4 & 15 & 18,084~~ & 1,498.1~~ & 32.908~~~~ & $3.6682 \times 10^{26}$ & +26.5645\\
7.6487 & 4 & 15 & 28,592~~ & 1,356.3~~ & 58.503~~~~ & $3.6034 \times 10^{26}$ & +26.5567\\
8.9234 & 4 & 15& 14,126~~ & 502.9~~ & 79.629~~~~ & $3.5272 \times 10^{26} $ & +26.5475\\
10.1983 & 4 & 10& 494,863~~ & 1,084.0~~ & 104.005~~~~ & $4.3893 \times 10^{27}$ & +27.6424 \\
12.1104 & 4 & 15 & 25,250~~ & 530.8~~ & 146.664~~~~ & $3.2431 \times 10^{26}$ & +26.5110\\
14.0226 & 4 & 15 &  28,259~~ & 501.2~~ & 196.636~~~~ & $2.8671 \times 10^{26} $ & +26.4575\\
15.9348 & 4 & 15 & 32,145~~ & 511.9~~ & 253.918~~~~ & $2.4729 \times 10^{26} $ & +26.3932 \\
17.8470 & 4 & 15 & 30,626~~ & 524.3~~ & 318.515~~~~  & $1.8339 \times 10^{26} $ & +26.2634\\
\hline

\hline
\end{tabular}
\label{table:nonlin}
\end{table*}

\begin{table*}
\caption{Technical specifications of jFFS iterations for $L=3.1869$~nm. $N_c, N_t$ and $N_s$ correspond to the number of starting configurations, trial trajectories and number of successful crossings, respectively. $t_{\text{succ}}$ and $t_{\text{fail}}$ are the average length of a successful and a failing trajectory, respectively. }
\centering
\begin{tabular}{c r r r r r r r c c}
\hline\hline`
$k$ & ~~~$\lambda_k$~~~ & $\lambda_{k+1}$~~ & $N_c$~~ & $N_t$~~~~ & ~~~$N_s$~~~ & $t_\text{succ}$~[fs]~~ & $t_\text{fail}$~[fs]~~ &$P(\lambda_{k+1}|\lambda_k)$ & $\log_{10}\left[P(\lambda_{k+1}|\lambda_k)\right]$  \\ [0.5ex] 
\hline
0 &  15~~~~ & 27~~~~ & 3,862~~ & 5,262,675 & 4,582~~ & 348.90~~~ &   163.86~~~ &0.00087&$-$3.0602 $\pm$ 0.1306 \\
1 &  27~~~~ & 42~~~~ & 4,582~~ & 747,880 & 4,022~~ &1,576.11~~~ &   551.33~~~	&0.00053&$-$3.2694 $\pm$ 0.1062\\
2 & 42~~~~ & 54~~~~ & 4,022~~ & 483,937 &3,385~~ &3,133.54~~~ &  2,014.42~~~	&0.00699&$-$2.1552 $\pm$ 0.0527\\
3 & 54~~~~ &  72~~~~ & 3,385~~ & 297,944 &3,302~~ &6,853.65~~~ &  6,788.42~~~	&0.01108&$-$1.9554 $\pm$ 0.0444\\
4 & 72~~~~ &  90~~~~ & 3,302~~ & 70,479 &3,397~~ &9,115.96~~~  & 15,930.51~~~	&0.04819&$-$1.3170 $\pm$ 0.0292\\
5 & 90~~~~ &  110~~~~ & 3,397~~ & 34,160 &3,356~~ &11,656.57~~~ &  28,121.92~~~	&0.09824&$-$1.0077 $\pm$ 0.0253\\
6 & 110~~~~ &  130~~~~ & 3,356~~ & 16,929 &4,085~~ &14,842.25~~~ &  44,146.85~~~	&0.24130&$-$0.6174 $\pm$ 0.0198\\
7 & 130~~~~ & 155~~~~ & 4,085~~ &  8,596  &3,493~~ &23,016.45~~~ &  65,494.74~~~	&0.40635&$-$0.3911 $\pm$ 0.0179\\
8 & 155~~~~ & 190~~~~ &  3,493~~ &  6,385 &4,409~~ &37,555.79~~~ &  97,772.46~~~ 	&0.69052& $-$0.1608 $\pm$ 0.0114\\
9 & 190~~~~ & 260~~~~ &  4,409~~ & 3,365 &3,213~~ &58,782.39~~~  & 144,448.03~~~	&0.95482& $-$0.0201 $\pm$ 0.0048\\[1ex]
\hline
\end{tabular}
\label{table:nonlin}
\end{table*}

\begin{table*}
\caption{Technical specifications of jFFS iterations for $L=3.3803$~nm. $N_c, N_t$ and $N_s$ correspond to the number of starting configurations, trial trajectories and number of successful crossings, respectively. $t_{\text{succ}}$ and $t_{\text{fail}}$ are the average length of a successful and a failing trajectory, respectively. }
\centering
\begin{tabular}{c r r r r r r r c c}
\hline\hline
$k$ & ~~~$\lambda_k$~~~ & $\lambda_{k+1}$~~ & $N_c$~~ & $N_t$~~~~ & ~~~$N_s$~~~ & $t_\text{succ}$~[fs]~~ & $t_\text{fail}$~[fs]~~ &$P(\lambda_{k+1}|\lambda_k)$ & $\log_{10}\left[P(\lambda_{k+1}|\lambda_k)\right]$  \\ [0.5ex] 
\hline
0 & 15~~~~ &25~~~~ &3,998~~ &1,837,072 &4,177~~ &	299.25~~~ &  163.76~~~ 	&0.00227&$-$2.6432 $\pm$ 0.0788 \\
1 & 25~~~~ &37~~~~ &4,177~~ &1,393,045 &4,120~~ &	1,398.65~~~  &   487.93~~~ 	&0.00295&$-$2.5291 $\pm$ 0.0577\\
2 & 37~~~~ &49~~~~ &4,120~~ &502,281 &4,172~~ &2,563.58~~~  &  2,076.14~~~ 		&0.00830&$-$2.0806 $\pm$ 0.0431\\
3 & 49~~~~ &62~~~~ &4,172~~ &276,404 &3,560~~ &3,643.53~~~  &   5,336.30~~~ 		&0.01287&$-$1.8901 $\pm$ 0.0386\\
4 & 62~~~~ &75~~~~ &3,560~~ &142,389 &3,528~~ & 4,867.64~~~  &   10,364.60~~~ 		&0.02477&$-$1.6059 $\pm$ 0.0359\\
5 & 75~~~~ &92~~~~ &3,528~~ &123,377 &3,590~~ &7,819.81~~~  &   17,973.12~~~ 		&0.02909&$-$1.5361 $\pm$ 0.0326\\
6 & 92~~~~ &110~~~~ &3,590~~ &37,772 &2,533~~ &10,628.12~~~  &   29,949.84~~~ 		&0.06706&$-$1.1735 $\pm$ 0.0304\\
7 & 110~~~~ &130~~~~ &2,533~~ &17,111 &2,559~~ &14,833.12~~~  &   45,741.27~~~ 		&0.14955&$-$0.8252 $\pm$ 0.0279\\
8 & 130~~~~ &150~~~~ &2,559~~ &7,193 &2,624~~ &19,067.95~~~  &   64,678.82~~~ 		&0.36479&$-$0.4379 $\pm$ 0.0220\\
9 & 150~~~~ &185~~~~ &2,624~~ &4,791 &2,570~~ & 45,373.35~~~  &   98,000.46~~~ 		&0.53642&$-$0.2705 $\pm$ 0.0183\\
10 & 185~~~~ &245~~~~ &2,570~~ &3,998 &3,544~~ &80,578.04~~~  &  161,973.55~~~ 		&0.88644&$-$0.0523 $\pm$ 0.0075\\[1ex]
\hline
\end{tabular}
\label{table:nonlin}
\end{table*}

\begin{table*}[ht]
\caption{Technical specifications of jFFS iterations for $L=3.6056$~nm. $N_c, N_t$ and $N_s$ correspond to the number of starting configurations, trial trajectories and number of successful crossings, respectively. $t_{\text{succ}}$ and $t_{\text{fail}}$ are the average length of a successful and a failing trajectory, respectively. }
\centering
\begin{tabular}{c r r r r r r r c c}
\hline\hline
$k$ & ~~~$\lambda_k$~~~ & $\lambda_{k+1}$~~ & $N_c$~~ & $N_t$~~~~ & ~~~$N_s$~~~ & $t_\text{succ}$~[fs]~~ & $t_\text{fail}$~[fs]~~ &$P(\lambda_{k+1}|\lambda_k)$ & $\log_{10}\left[P(\lambda_{k+1}|\lambda_k)\right]$  \\ [0.5ex] 
\hline
0 & 15~~~~ &25~~~~ &3,594~~ &2,976,195&6,108~~ &	274.53~~~   & 171.38~~~  &	0.00205&$-$2.6878 $\pm$ 0.0710 \\
1 & 25~~~~ &37~~~~ &6,108~~ &5,201,885&4,303~~ &	853.88~~~  &  412.43~~~ &	0.00082&$-$3.0824 $\pm$ 0.0895\\
2 & 37~~~~ &47~~~~ &4,303~~ &535,339&4,765~~ &2,130.50~~~   & 1,469.02~~~ 	&	0.00890&$-$2.0506 $\pm$ 0.0456\\
3 & 47~~~~ &59~~~~ &4,765~~ &221,557&4,336~~ &3,801.90~~~   & 5,337.76~~~ 	&	0.01957&$-$1.7084 $\pm$ 0.0355\\
4 & 59~~~~ &76~~~~ &4,336~~ &207,661&3,162~~ &6,810.31~~~  & 11,243.84~~~ 	&	0.01522&$-$1.8174 $\pm$ 0.0321\\
5 & 76~~~~ &91~~~~ &3,163~~ &65,872&3,310~~ &6,613.14~~~   & 19,941.93~~~ 	&	0.05024&$-$1.2989 $\pm$ 0.0302\\
6 & 91~~~~ &107~~~~ &3,310~~ &41,855&3,117~~ &8,450.46~~~   & 29,321.15~~~ 	&	0.07447&$-$1.1280 $\pm$ 0.0272\\
7 & 107~~~~ &124~~~~ &3,117~~ &27,281&3,098~~ &	9,796.81~~~  &  40,144.86~~~ &	0.11355&$-$0.9448 $\pm$ 0.0260\\
8 & 124~~~~ &144~~~~ &3,098~~ &17,344&2,636~~ &13,867.81~~~ &  53,683.35~~~ 	&	0.15198&$-$0.8182 $\pm$ 0.0262\\
9 & 144~~~~ &165~~~~ &2,636~~ &9,676&2,714~~ &17,879.50~~~  &  71,893.90~~~ &		0.28048&$-$0.5521 $\pm$ 0.0234\\
10 & 165~~~~ &200~~~~ &2,714~~ &6,432&2,567~~ &40,717.67~~~   & 100,214.98~~~ &		0.39909&$-$0.3989 $\pm$ 0.0215\\
11 & 200~~~~ &260~~~~ &2,567~~ &3,263&2,609~~ &70,161.00~~~   & 151,523.06~~~ &		0.79957&$-$0.0971 $\pm$ 0.0117\\[1ex]
\hline
\end{tabular}
\label{table:nonlin}
\end{table*}

\begin{table*}[ht]
\caption{Technical specifications of jFFS iterations for $L=3.8243$~nm. $N_c, N_t$ and $N_s$ correspond to the number of starting configurations, trial trajectories and number of successful crossings, respectively. $t_{\text{succ}}$ and $t_{\text{fail}}$ are the average length of a successful and a failing trajectory, respectively. }
\centering
\begin{tabular}{c r r r r r r r c c}
\hline\hline
$k$ & ~~~$\lambda_k$~~~ & $\lambda_{k+1}$~~ & $N_c$~~ & $N_t$~~~~ & ~~~$N_s$~~~ & $t_\text{succ}$~[fs]~~ & $t_\text{fail}$~[fs]~~ &$P(\lambda_{k+1}|\lambda_k)$ & $\log_{10}\left[P(\lambda_{k+1}|\lambda_k)\right]$  \\ [0.5ex] 
\hline
0 & 15~~~~ &27~~~~ &6,298~~ &7,812,913&5,765~~ & 408.85~~~  & 168.69~~~  & 0.00073& $-$3.1320 $\pm$ 0.0704 \\
1 & 27~~~~ &38~~~~ &5,765~~ &1,571,503&4,569~~ & 1,007.51~~~  & 564.95~~~  & 0.00290& $-$2.5365 $\pm$ 0.0573\\
2 & 38~~~~ &54~~~~ &4,569~~ &1,835,735&3,071~~ & 2,921.67~~~  &  1,836.92~~~ 		&0.00167& $-$2.7765 $\pm$ 0.0736\\
3 & 54~~~~ &68~~~~ &3,071~~ &366,787&3,602~~ & 4,445.29~~~   & 5,826.90~~~ 	&0.00982&$-$2.0079 $\pm$ 0.0471\\
4 & 68~~~~ &82~~~~ &3,602~~ &95,817&3,289~~ & 5,647.28~~~   & 13,254.90~~~ 		&0.03432&$-$1.4644 $\pm$ 0.0325\\
5 & 82~~~~ &101~~~~ &3,289~~ &10,4312&3,211~~ & 9,267.57~~~  &  23,094.28~~~ &0.03078&$-$1.5117 $\pm$ 0.0301\\
6 & 101~~~~ &118~~~~ &3,211~~ &34,990&2,900~~ & 9,668.87~~~   & 35,570.67~~~  &0.08288&$-$1.0815 $\pm$ 0.0271\\
7 & 118~~~~ &138~~~~ &2,900~~ &28,240&2,799~~ &	12,300.06~~~   & 48,794.57~~~ 	 &0.09911&$-$1.0039 $\pm$ 0.0270\\
8 & 138~~~~ &160~~~~ &2,799~~ &20,543&2,971~~ &	16,130.64~~~   & 65,101.95~~~ 	 &0.14462&$-$0.8398 $\pm$ 0.0245\\
9 & 160~~~~ &185~~~~ &2,971~~ &12,890&2,882~~ &	22,636.19~~~  &  86,123.50~~~ 	 &0.22358&$-$0.6506 $\pm$ 0.0235\\
10 & 185~~~~ &210~~~~ &2,882~~ &6,919&3,405~~ & 26,813.50~~~   & 112,919.88~~~ 	 &0.49212&$-$0.3079 $\pm$ 0.0172\\
11 & 210~~~~ &245~~~~ &3,405~~ &5,624&4,188~~ & 40,801.99~~~   & 146,352.29~~~ 	 &0.74466&$-$0.1280 $\pm$ 0.0106\\
12 & 245~~~~ &300~~~~ &4,188~~ &3,584&3,411~~ & 47,489.89~~~   & 187,497.11~~~  &0.95172&$-$0.0215 $\pm$ 0.0050\\[1ex]
\hline
\end{tabular}
\label{table:nonlin}
\end{table*}

\begin{table*}
\caption{Technical specifications of jFFS iterations for $L=4.0563$~nm. $N_c, N_t$ and $N_s$ correspond to the number of starting configurations, trial trajectories and number of successful crossings, respectively. $t_{\text{succ}}$ and $t_{\text{fail}}$ are the average length of a successful and a failing trajectory, respectively. }
\centering
\begin{tabular}{c r r r r r r r c c}
\hline\hline
$k$ & ~~~$\lambda_k$~~~ & $\lambda_{k+1}$~~ & $N_c$~~ & $N_t$~~~~ & ~~~$N_s$~~~ & $t_\text{succ}$~[fs]~~ & $t_\text{fail}$~[fs]~~ &$P(\lambda_{k+1}|\lambda_k)$ & $\log_{10}\left[P(\lambda_{k+1}|\lambda_k)\right]$  \\ [0.5ex] 
\hline
0 & 15~~~~ &24~~~~ &4,379~~ &1,172,372&4,609~~ &	262.21~~~ &   168.81~~~ 	&0.00393&$-$2.4055 $\pm$ 0.0634 \\
1 & 24~~~~ &35~~~~ &4,609~~ &860,778&5,081~~ &1,238.23~~~   & 463.54~~~ 	&0.00590&$-$2.2289 $\pm$ 0.0577\\
2 & 35~~~~ &46~~~~ &5,081~~ &336,147&4,489~~ &2,210.95~~~   & 2,215.68~~~ 		&0.01335& $-$1.8744 $\pm$ 0.0310\\
3 & 46~~~~ &60~~~~ & 4,489~~ &460,325&3,638~~ &3,828.01~~~  &  4,659.00~~~ 	&0.00790& $-$2.1022 $\pm$ 0.0443\\
4 & 60~~~~ &78~~~~ &3,638~~ &392,244&3,024~~ &6,548.95~~~   & 9,963.22~~~ 		&0.00770&$-$2.1130 $\pm$ 0.0407 \\
5 & 78~~~~ &98~~~~ & 3,024~~ &138,608&2,558~~ &9,459.56~~~  &  20,032.57~~~ 		&0.01841& $-$1.7349 $\pm$ 0.0362\\
6 & 98~~~~ &121~~~~ &2,558~~ &81,358&2,534~~ &12,938.75~~~  &  33,396.95~~~ 		&0.03114&$-$1.5066 $\pm$ 0.0337\\
7 & 121~~~~ &150~~~~ &2,534~~ &65,364&2,513~~ &19,865.04~~~   & 51,708.82~~~ 		&0.03844&$-$1.4151 $\pm$ 0.0297\\
8 & 150~~~~ &180~~~~ &2,513~~ &26,462&2,567~~ &24,198.86~~~   & 77,521.55~~~ 		&0.09700&$-$1.0132 $\pm$ 0.0269\\
9 & 180~~~~ &210~~~~ &2,567~~ &15,150&3,161~~ &	29,779.70~~~  &  110,267.50~~~ 	&0.20864&$-$0.6806 $\pm$ 0.0229\\
10 & 210~~~~ &250~~~~ &3,161~~ &10,841&4,013~~ &	49,575.38~~~  & 153,581.05~~~ 	&0.37016&$-$0.4316 $\pm$ 0.0180\\
11 & 250~~~~ &300~~~~ &4,013~~ &5,532&4,021~~ &	61,627.03~~~ &  214,156.25~~~ 	&0.72686& $-$0.1385 $\pm$ 0.0112\\[1ex]
\hline
\end{tabular}
\label{table:nonlin}
\end{table*}

\begin{table*}
\caption{Technical specifications of jFFS iterations for $L=4.2817$~nm. $N_c, N_t$ and $N_s$ correspond to the number of starting configurations, trial trajectories and number of successful crossings, respectively. $t_{\text{succ}}$ and $t_{\text{fail}}$ are the average length of a successful and a failing trajectory, respectively. }
\centering
\begin{tabular}{c r r r r r r r c c}
\hline\hline
$k$ & ~~~$\lambda_k$~~~ & $\lambda_{k+1}$~~ & $N_c$~~ & $N_t$~~~~ & ~~~$N_s$~~~ & $t_\text{succ}$~[fs]~~ & $t_\text{fail}$~[fs]~~ &$P(\lambda_{k+1}|\lambda_k)$ & $\log_{10}\left[P(\lambda_{k+1}|\lambda_k)\right]$  \\ [0.5ex] 
\hline
0 & 15~~~~  & 29~~~~  & 12,031~~   & 20,584,990 & 5,430~~   & 582.21~~~ & 174.40~~~ & 0.000264& $-$3.5788 $\pm$ 0.1117 \\ 
1 & 29~~~~  & 41~~~~  & 5,430~~   & 1,644,639 & 6,577~~   & 1,696.83~~~  & 800.31~~~  & 0.003999 & $-$2.3980 $\pm$ 0.0409\\
2 & 41~~~~  & 53~~~~  & 6,577~~   & 645,282 & 6,133~~   &  2,794.47~~~  &  2,807.61~~~ 	& 0.009504 & $-$2.0221 $\pm$ 0.0325\\
3 & 53~~~~  & 67~~~~  & 6,133~~   & 431,240 & 5,821~~   & 4,448.33~~~   & 6,634.67~~~  & 0.013498 & $-$1.8697 $\pm$ 0.0309\\
4 & 67~~~~  & 83~~~~  & 5,821~~   & 254,692 & 5,577~~   & 6,325.95~~~  & 13,293.85~~~  & 0.021897 & $-$1.6596 $\pm$ 0.0265\\
5 & 83~~~~  & 100~~~~  & 5,577~~   & 129,181 & 5,549~~   & 7,928.05~~~  &  23,038.20~~~  & 0.042955& $-$1.3670 $\pm$ 0.0230\\
6 & 100~~~~  & 120~~~~  & 5,549~~   & 101,337 & 5,578~~   & 11,288.84~~~  & 35,146.29~~~  & 0.054927& $-$1.2602 $\pm$ 0.0207\\
7 & 120~~~~  & 150~~~~  & 5,578~~   & 14,0347 & 5,172~~   & 20,342.82~~~   & 51,683.75~~~  & 0.036851 & $-$1.4335 $\pm$ 0.0207\\
8 & 150~~~~  & 200~~~~  & 5,172~~   & 17,2972 & 5,195~~   & 44,856.77~~~  & 82,160.85~~~  & 0.030034 &  $-$1.5224 $\pm$ 0.0198\\
9 & 200~~~~  & 270~~~~  & 5,195~~   & 41,575 & 5,348~~   & 87,744.02~~~  &  149,928.90~~~ & 0.128635& $-$0.8906 $\pm$ 0.0183\\
10 & 270~~~~  & 370~~~~  & 5,348~~   & 12,000 & 8,693~~   & 109,688.79~~~  & 265,100.61~~~  & 0.724417 & $-$0.1400 $\pm$ 0.0078\\
\hline
\end{tabular}
\label{table:nonlin}
\end{table*}

\begin{table*}[ht]
\caption{Technical specifications of jFFS iterations for $L=4.4617$~nm. $N_c, N_t$ and $N_s$ correspond to the number of starting configurations, trial trajectories and number of successful crossings, respectively. $t_{\text{succ}}$ and $t_{\text{fail}}$ are the average length of a successful and a failing trajectory, respectively. }
\centering
\begin{tabular}{c r r r r r r r c c}
\hline\hline
$k$ & ~~~$\lambda_k$~~~ & $\lambda_{k+1}$~~ & $N_c$~~ & $N_t$~~~~ & ~~~$N_s$~~~ & $t_\text{succ}$~[fs]~~ & $t_\text{fail}$~[fs]~~ &$P(\lambda_{k+1}|\lambda_k)$ & $\log_{10}\left[P(\lambda_{k+1}|\lambda_k)\right]$  \\ [0.5ex] 
\hline
0 & 15~~~~ &27~~~~ &12,511~~ &6,371,497&6,107~~ &382.14~~~   &  175.13~~~ &0.00096& $-$3.0184$\pm$ 0.0764 \\
1 & 27~~~~ &42~~~~ &6,107~~ &4,531,030&5,396~~ &2,030.29~~~  &   615.76~~~ 		&0.00119& $-$2.9241 $\pm$ 0.0622\\
2 & 42~~~~ &58~~~~ &5,396~~ &1,356,236&5,588~~ &	4,583.24~~~   & 3,092.55~~~ 	&0.00412&$-$2.3851 $\pm$ 0.0435\\
3 & 58~~~~ &76~~~~ &5,588~~ &565,654&5,483~~ &	7,422.68~~~   & 9,391.14~~~ 	&0.00969&$-$2.0135 $\pm$ 0.0305\\
4 & 76~~~~ &96~~~~ &5,483~~ &268,813&5,944~~ &	9,787.80~~~  &  20,063.14~~~ 	&0.02211& $-$1.6554 $\pm$ 0.0248\\
5 & 96~~~~ &120~~~~ &5,944~~ &208,619&6,446~~ &	13,829.84~~~  &  33,829.64~~~ 	&0.03090&$-$1.5101 $\pm$ 0.0204\\
6 & 120~~~~ &150~~~~ &6,446~~ &142,225&5,755~~ & 20,048.19~~~ &  52,135.56~~~ 		&0.04046&$-$1.3929 $\pm$ 0.0191\\
7 & 150~~~~ &190~~~~ &5,755~~ &110,618&5,562~~ & 33,642.70~~~  & 79,794.80~~~ 		&0.05028&$-$1.2986 $\pm$ 0.0183\\
8 & 190~~~~ &260~~~~ &5,562~~ &83,488&5,357~~ &	83,202.89~~~   & 132,550.75~~~ 	&0.06416&$-$1.1927 $\pm$ 0.0178\\
9 & 260~~~~ &350~~~~ &5,357~~ &11,000&4,666~~ &	138,702.00~~~  &  257,638.69~~~ 	&0.42418&$-$0.3724 $\pm$ 0.0149\\
10 & 350~~~~ &450~~~~ &4,666~~ &3,800&3,640~~ &98,515.11~~~ 	 &391,049.31~~~ 	&0.95789&$-$0.0187 $\pm$ 0.0044\\[1ex]
\hline
\end{tabular}
\label{table:nonlin}
\end{table*}

\begin{table*}
\caption{Technical specifications of jFFS iterations for $L=5.0991$~nm. $N_c, N_t$ and $N_s$ correspond to the number of starting configurations, trial trajectories and number of successful crossings, respectively. $t_{\text{succ}}$ and $t_{\text{fail}}$ are the average length of a successful and a failing trajectory, respectively. }
\centering
\begin{tabular}{c r r r r r r r c c}
\hline\hline
$k$ & ~~~$\lambda_k$~~~ & $\lambda_{k+1}$~~ & $N_c$~~ & $N_t$~~~~ & ~~~$N_s$~~~ & $t_\text{succ}$~[fs]~~ & $t_\text{fail}$~[fs]~~ &$P(\lambda_{k+1}|\lambda_k)$ & $\log_{10}\left[P(\lambda_{k+1}|\lambda_k)\right]$  \\ [0.5ex] 
\hline
0 & 15~~~~ &28~~~~ &6,433~~ &6,969,031&4,341~~ &	550.66~~~ & 185.46~~~ 	&0.00062& $-$3.2056 $\pm$ 0.1383 \\
1 & 28~~~~ &41~~~~ &4,341~~ &1,417,200&4,814~~ &	1,862.32~~~  &  844.85~~~ 	&0.00339&$-$2.4689 $\pm$ 0.0565 \\
2 & 41~~~~ &55~~~~ &4,814~~ &559,509&3,251~~ &3,349.57~~~ &  2,911.58~~~ 		&0.00581&$-$2.2358 $\pm$ 0.0436\\
3 & 55~~~~ &69~~~~ &3,251~~ &209,273&3,263~~ & 4,343.49~~~  &  7,089.15~~~ 		&0.01559&$-$1.8071 $\pm$ 0.0399\\
4 & 69~~~~ &87~~~~ &3,263~~ &216,058&3,116~~ &7,376.38~~~  &  13,959.59~~~ 		&0.01442&$-$1.8410 $\pm$ 0.0372\\
5 & 87~~~~ &105~~~~ &3,116~~ &78,366&3,193~~ &8,326.87~~~  &  25,177.98~~~ 		&0.04074&$-$1.3899 $\pm$ 0.0305\\
6 & 105~~~~ &123~~~~ &3,193~~ &35,215&2,760~~ &	9,644.38~~~  &  37,653.95~~~ 	&0.07837&$-$1.1058 $\pm$ 0.0278\\
7 & 123~~~~ &145~~~~ &2,760~~ &32,312&2,713~~ &	14,312.30~~~  & 51,556.38~~~ &0.08396&$-$1.0759 $\pm$ 0.0273 \\
8 & 145~~~~ &166~~~~ &2,713~~ &17,329&2,686~~ &	15,582.50~~~  & 69,927.54~~~ 	&0.15500&$-$0.8097 $\pm$ 0.0258\\
9 & 166~~~~ &188~~~~ &2,686~~ &12,192&2,635~~ &	 18,347.65~~~  &  89,835.75~~~ 	&0.21612&$-$0.6653 $\pm$ 0.0241\\
10 & 188~~~~ &212~~~~ &2,635~~ & 9,318&2,568~~ &24,175.91~~~  &  114,499.99~~~ 		&0.27559&$-$0.5597 $\pm$ 0.0231\\
11 & 212~~~~ &240~~~~ &2,568~~ &8,497&3,107~~ &31,993.68~~~  &  145,381.67~~~ 		&0.36565&$-$0.4369 $\pm$ 0.0197\\
12 & 240~~~~ &270~~~~ &3,107~~ &6,363&3,071~~ & 41,096.51~~~  &  186,614.25~~~ 		&0.48263&$-$0.3164 $\pm$ 0.0175\\
13 & 270~~~~ &310~~~~ &3,071~~ &4,190&2,609~~ & 62,110.51~~~  &  244,627.99~~~ 		&0.62267&$-$0.2057 $\pm$ 0.0158\\
14 & 310~~~~ &370~~~~ &2,609~~ &3,305&2,628~~ &99,413.37~~~   & 336,098.17~~~ 	&0.79515& $-$0.0995 $\pm$ 0.0117\\[1ex]
\hline
\end{tabular}
\label{table:nonlin}
\end{table*}

\begin{table*}[ht]
\caption{Technical specifications of jFFS iterations for $L=5.7365$~nm. $N_c, N_t$ and $N_s$ correspond to the number of starting configurations, trial trajectories and number of successful crossings, respectively. $t_{\text{succ}}$ and $t_{\text{fail}}$ are the average length of a successful and a failing trajectory, respectively. }
\centering
\begin{tabular}{c r r r r r r r c c}
\hline\hline
$k$ & ~~~$\lambda_k$~~~ & $\lambda_{k+1}$~~ & $N_c$~~ & $N_t$~~~~ & ~~~$N_s$~~~ & $t_\text{succ}$~[fs]~~ & $t_\text{fail}$~[fs]~~ &$P(\lambda_{k+1}|\lambda_k)$ & $\log_{10}\left[P(\lambda_{k+1}|\lambda_k)\right]$  \\ [0.5ex] 
\hline
0 & 15~~~~ &27~~~~ &11,764~~ &4,639,199&4,380~~ &460.80~~~  &   195.41~~~  &0.00094&$-$3.0250 $\pm$ 0.0708 \\
1 & 27~~~~ &40~~~~ &4,380~~ &1,096,800&4,298~~ &1,734.78~~~  &   713.80~~~  &0.00391&$-$2.4069 $\pm$ 0.0579\\
2 & 40~~~~ &58~~~~ &4,298~~ &2,099,982&3,035~~ &	4,143.57~~~  &   3,115.44~~~  &0.00144&$-$2.8401 $\pm$ 0.0630\\
3 & 58~~~~ &75~~~~ &3,035~~ &508,904&3,049~~ &5,338.93~~~  &   7,796.36~~~  &0.00599&$-$2.2225 $\pm$ 0.0555\\
4 & 75~~~~ &90~~~~ &3,049~~ &82,307&3,085~~ &5,723.98~~~  &   16,371.97~~~  &0.03748&$-$1.4262 $\pm$ 0.0338\\
5 & 90~~~~ &110~~~~ &3,085~~ &196,003&6,307~~ &	9,514.35~~~  & 26,690.49~~~ &0.03217&$-$1.4924 $\pm$ 0.0264\\
6 & 110~~~~ &130~~~~ &6,307~~ &41,446&2,587~~ &11,420.49~~~  &   40,326.62~~~  &0.06241&$-$1.2047 $\pm$ 0.0275\\
7 & 130~~~~ &150~~~~ &2,587~~ &23,157&2,652~~ &	13,022.31~~~  &   55,903.65~~~ &0.11452&$-$0.9411 $\pm$ 0.0273\\
8 & 150~~~~ &170~~~~ &2,652~~ &14,772&2,638~~ &	15,201.87~~~  &   72,975.46~~~  &0.17858&$-$0.7482 $\pm$ 0.0251\\
9 & 170~~~~ &190~~~~ &2,638~~ &10,488&2,727~~ &16,695.66~~~  &   91,556.28~~~ &0.26001& $-$0.5850 $\pm$ 0.0233\\
10 & 190~~~~ &214~~~~ &2,727~~ &8,980&2,562~~ &	 24,971.90~~~  &   113,922.85~~~  &0.28530&$-$0.5447 $\pm$ 0.0232\\
11 & 214~~~~ &240~~~~ &2,562~~ &7,091&2,661~~ &28,877.21~~~  &   144,178.29~~~ 	&0.37526&$-$0.4257 $\pm$ 0.0212\\
12 & 240~~~~ &270~~~~ &2,661~~ &5,694&2,709~~ &40,618.95~~~   & 185,424.21~~~ 	&0.47576&$-$0.3226 $\pm$ 0.0188\\
13 & 270~~~~ &315~~~~ &2,709~~ &4,459&2,530~~ &	71,045.71~~~  &   249,814.76~~~  &0.56739& $-$0.2461$\pm$ 0.0174\\
14 & 315~~~~ &380~~~~ &2,530~~ &3,174&2,541~~ &106,053.45~~~  &   339,170.14~~~ 	&0.80056&$-$0.0966 $\pm$ 0.0117\\[1ex]
\hline
\end{tabular}
\label{table:nonlin}
\end{table*}

\begin{table*}
\caption{Technical specifications of jFFS iterations for $L=7.6487$~nm. $N_c, N_t$ and $N_s$ correspond to the number of starting configurations, trial trajectories and number of successful crossings, respectively. $t_{\text{succ}}$ and $t_{\text{fail}}$ are the average length of a successful and a failing trajectory, respectively. }
\centering
\begin{tabular}{c r r r r r r r c c}
\hline\hline
$k$ & ~~~$\lambda_k$~~~ & $\lambda_{k+1}$~~ & $N_c$~~ & $N_t$~~~~ & ~~~$N_s$~~~ & $t_\text{succ}$~[fs]~~ & $t_\text{fail}$~[fs]~~ &$P(\lambda_{k+1}|\lambda_k)$ & $\log_{10}\left[P(\lambda_{k+1}|\lambda_k)\right]$  \\ [0.5ex] 
\hline
0 & 15~~~~  & 29~~~~  & 17,854~~  & 10,991,089 & 4,375~~  & 815.17~~~ &  242.20~~~  & 0.000398 & $-$3.4000 $\pm$ 0.0687 \\
1 & 29~~~~  & 48~~~~  & 4,375~~  & 12,885,977 & 4,597~~  & 3,852.66~~~  & 1,098.29~~~  & 0.000357 & $-$3.4476 $\pm$ 0.0999 \\
2 & 48~~~~  & 64~~~~  & 4,597~~  & 768,264& 4,659~~ & 5,157.04~~~  &   5,388.14~~~  & 0.006064 &  $-$2.2172 $\pm$ 0.0394 \\
3 & 64~~~~  & 78~~~~  & 4,659~~  & 169,391 & 4,815~~  & 5,305.28~~~  & 12,697.17~~~ & 0.028425 & $-$1.5463 $\pm$ 0.0289 \\
4 & 78~~~~  & 94~~~~  & 4,815~~  & 121,463 & 4,741~~  & 7,160.93~~~  & 20,741.46~~~  & 0.039032 & $-$1.4086 $\pm$ 0.0256\\
5 & 94~~~~  & 115~~~~  & 4,741~~  & 111,846 & 4,219~~  & 11,039.72~~~  & 31,954.20~~~  & 0.037722 & $-$1.4234 $\pm$ 0.0250\\
6 & 115~~~~  & 138~~~~  & 4,219~~  & 74,624 & 4,280~~  & 14,240.34~~~  & 47,143.39~~~  & 0.057354 & $-$1.2414 $\pm$ 0.0224\\
7 & 138~~~~  & 165~~~~  & 4,280~~  & 54,225 & 4,205~~  & 19,952.40~~~  & 66,563.31~~~  & 0.077547 & $-$1.1104 $\pm$ 0.0213\\
8 & 165~~~~  & 200~~~~  & 4,205~~  & 43,628 & 4,097~~  & 31,347.47~~~  & 94,080.43~~~  & 0.093907 & $-$1.0273 $\pm$ 0.0207\\
9 & 200~~~~  & 245~~~~  & 4,097~~  & 30,318 & 4,695~~  & 49,839.47~~~  &139,413.30~~~  & 0.154858& $-$0.8101 $\pm$ 0.0182\\
10 & 245~~~~  & 310~~~~  & 4,695~~  & 13,342 & 4,082~~  & 102,984.76~~~  & 222,090.20~~~  & 0.305951 & $-$0.5143 $\pm$ 0.0177\\
11 & 310~~~~  & 390~~~~  & 4,082~~  & 8,966 & 6,681~~  & 131,355.52~~~  & 340,236.93~~~  & 0.745148 & $-$0.1278 $\pm$ 0.0084\\
\hline
\end{tabular}
\label{table:nonlin}
\end{table*}

\begin{table*}[ht]
\caption{Technical specifications of jFFS iterations for $L=8.9234$~nm. $N_c, N_t$ and $N_s$ correspond to the number of starting configurations, trial trajectories and number of successful crossings, respectively. $t_{\text{succ}}$ and $t_{\text{fail}}$ are the average length of a successful and a failing trajectory, respectively. }
\centering
\begin{tabular}{c r r r r r r r c c}
\hline\hline
$k$ & ~~~$\lambda_k$~~~ & $\lambda_{k+1}$~~ & $N_c$~~ & $N_t$~~~~ & ~~~$N_s$~~~ & $t_\text{succ}$~[fs]~~ & $t_\text{fail}$~[fs]~~ &$P(\lambda_{k+1}|\lambda_k)$ & $\log_{10}\left[P(\lambda_{k+1}|\lambda_k)\right]$  \\ [0.5ex] 
\hline
0 & 15~~~~  & 27~~~~  & 8,667~~   & 3,400,311 & 4,612~~   & 597.83~~~    & 293.26~~~  & 0.001356 & $-2.8676\pm0.0644$ \\
1 & 27~~~~  & 45~~~~  & 4,612~~   & 19,580,158 & 4,072~~   & 3,143.53~~~   & 902.68~~~   & 0.000208 & $-3.6820\pm0.0771$ \\
2 & 45~~~~  & 58~~~~  & 4,070~~   & 429,743 & 4,036~~   & 3,719.91~~~    & 4,090.26~~~  & 0.009392 & $-2.0273\pm0.0434$ \\
3 & 58~~~~  & 72~~~~  & 4,036~~   & 131,975 & 3,315~~   & 5,257.47~~~   & 10,506.77~~~   &0.025118 & $-1.6000\pm0.0336$ \\
4 & 72~~~~  & 86~~~~  & 3,315~~   & 70,950 & 3,412~~   & 6,485.87~~~    & 18,741.19~~~  &		 0.048090 & $-1.3179\pm0.0281$ \\
5 & 86~~~~  & 100~~~~  & 3,412~~   & 35,670 & 3,114~~   &	7,099.87~~~    & 27,767.27~~~  &	 0.087300 & $-1.0590\pm0.0263$\\
6 & 100~~~~  & 116~~~~  & 3,114~~   & 33,244 & 3,271~~   &9,076.30~~~   &  37,231.01~~~  &		 0.098394 & $-1.0070\pm0.0256$\\
7 & 116~~~~  & 135~~~~  & 3,271~~   & 30,562 & 3,146~~   &11,940.06~~~    & 49,513.03~~~  	&	 0.102938 & $-0.9874\pm0.0248$\\
8 & 135~~~~  & 165~~~~  & 3,146~~   & 49,911 & 3,041~~   &22,538.74~~~   &  67,070.13~~~  	&	 0.060928 & $-1.2152\pm0.0251$\\
9 & 165~~~~  & 195~~~~  & 3,041~~   & 22,426 & 3,087~~   &26,714.90~~~  &  96,031.36~~~  	&	 0.137653 & $-0.8612\pm0.0234$ \\
10 & 195~~~~  & 230~~~~  & 3,087~~   & 14,243 & 3,076~~   &39,364.67~~~  &  133,406.20~~~  	&	 0.215966 & $-0.6656\pm0.0219$\\
11 & 230~~~~  & 270~~~~  & 3,076~~   & 19,687 & 6,839~~   & 55,655.23~~~   &  187,457.20~~~  	&	 0.347387 & $-0.4592\pm0.0135$\\
12 & 270~~~~  & 340~~~~  & 6,839~~   & 10,095 & 5,102~~   &118,452.39~~~   &  278,317.47~~~  	&	 0.505399 & $-0.2964\pm0.0131$\\
\hline
\end{tabular}
\label{table:nonlin}
\end{table*}


\begin{table*}
\caption{Technical specifications of jFFS iterations for $L=10.1983$~nm. $N_c, N_t$ and $N_s$ correspond to the number of starting configurations, trial trajectories and number of successful crossings, respectively. $t_{\text{succ}}$ and $t_{\text{fail}}$ are the average length of a successful and a failing trajectory, respectively. }
\centering
\begin{tabular}{c r r r r r r r c c}
\hline\hline
$k$ & ~~~$\lambda_k$~~~ & $\lambda_{k+1}$~~ & $N_c$~~ & $N_t$~~~~ & ~~~$N_s$~~~ & $t_\text{succ}$~[fs]~~ & $t_\text{fail}$~[fs]~~ &$P(\lambda_{k+1}|\lambda_k)$ & $\log_{10}\left[P(\lambda_{k+1}|\lambda_k)\right]$  \\ [0.5ex] 
\hline
0 & 10~~~~  &24~~~~  &87,274~~ &8,126,536&4,080~~ &771.46~~~   &  211.60~~~  		&0.00050& $-$3.2992 $\pm$ 0.0407  \\
1 & 24~~~~  &38~~~~  &4,080~~ &1,290,378&4,079~~ &	1,861.66~~~   &  1,068.34~~~  	&0.00316&$-$2.5002 $\pm$ 0.0685 \\
2 & 38~~~~  &52~~~~  &4,079~~ &613,783&4,050~~ &3,127.64~~~   &  3,280.94~~~  		&0.00659&$-$2.1806 $\pm$ 0.0435 \\
3 & 52~~~~  &70~~~~  &4,050~~ &541,810&2,516~~ &5,152.44~~~    & 7,401.30~~~  		&0.00464&$-$2.3331 $\pm$ 0.0492\\
4 & 70~~~~  &85~~~~  &2,516~~ &101,535&2,568~~ & 5,390.59~~~    & 15,181.82~~~  		&0.02529&$-$1.5970 $\pm$ 0.0415 \\
5 & 85~~~~  &100~~~~  &2,568~~ &45,700&2,553~~ &6,674.68~~~    & 24,413.33~~~  	&0.05586&$-$1.2529 $\pm$ 0.0329\\
6 & 100~~~~  &115~~~~   &2,553~~ &27,320&2,586~~ &7,632.50~~~   &  35,355.25~~~  		&0.09465&$-$1.0239 $\pm$ 0.0300\\
7 & 115~~~~  &130~~~~  &2,586~~ &17,831&2,640~~ &	8,816.08~~~   &  46,183.98~~~  	&0.14805&$-$0.8296 $\pm$ 0.0268  \\
8 & 130~~~~  &145~~~~  &2,640~~ &13,342&2,586~~ &10,071.11~~~   &  57,952.70~~~  		&0.19382&$-$0.7126 $\pm$ 0.0253 \\
9 & 145~~~~  &160~~~~  &2,586~~ &9,922&2,603~~ &10,777.74~~~  &  70,005.77~~~  	&0.26234& $-$0.5811 $\pm$ 0.0239 \\
10 & 160~~~~  &175~~~~  &2,603~~ &7,771&2,574~~ &11,545.94~~~   &  83,732.70~~~  		&0.33123&$-$0.4799 $\pm$ 0.0226 \\
11 & 175~~~~  &195~~~~  &2,574~~ &9,024&2,590~~ &18,453.66~~~  &  100,354.47~~~  	&0.28701&$-$0.5421 $\pm$ 0.0237 \\
12 & 195~~~~  &215~~~~  &2,590~~ &6,855&2,596~~ &20,026.02~~~   &  123,243.89~~~  	&0.37870&$-$0.4217 $\pm$ 0.0212 \\
13 & 215~~~~  &240~~~~  &2,596~~ &6,130&2,557~~ &28,910.39~~~   &  149,493.89~~~  		&0.41712&$-$0.3797 $\pm$ 0.0206 \\
14 & 240~~~~  &280~~~~  &2,557~~ &6,237&2,534~~ &56,995.43~~~   &  200,884.05~~~  	&0.40628&$-$0.3912 $\pm$ 0.0208 \\
15 & 280~~~~  &320~~~~  &2,534~~ &3,728&2,566~~ &62,187.21~~~   &  260,473.60~~~  		&0.68830&$-$0.1622 $\pm$ 0.0146 \\ 
16 & 320~~~~  &370~~~~  &2,566~~ &3,558&3,055~~ &	80,361.13~~~    & 331,816.73~~~  	&0.85862&$-$0.0662 $\pm$ 0.0089\\[1ex]
\hline
\end{tabular}
\label{table:nonlin}
\end{table*}

\begin{table*}[ht]
\caption{Technical specifications of jFFS iterations for $L=12.1104$~nm. $N_c, N_t$ and $N_s$ correspond to the number of starting configurations, trial trajectories and number of successful crossings, respectively. $t_{\text{succ}}$ and $t_{\text{fail}}$ are the average length of a successful and a failing trajectory, respectively. }
\centering
\begin{tabular}{c r r r r r r r c c}
\hline\hline
$k$ & ~~~$\lambda_k$~~~ & $\lambda_{k+1}$~~ & $N_c$~~ & $N_t$~~~~ & ~~~$N_s$~~~ & $t_\text{succ}$~[fs]~~ & $t_\text{fail}$~[fs]~~ &$P(\lambda_{k+1}|\lambda_k)$ & $\log_{10}\left[P(\lambda_{k+1}|\lambda_k)\right]$  \\ [0.5ex] 
\hline
0 & 15~~~~ &28~~~~ &14,992~~ &3,385,008&4031~~ &1,215.81~~~  &  583.73~~~ &0.00119&$-$2.9241 $\pm$ 0.0544 \\
1 & 28~~~~ &45~~~~ &4,031~~ &4,128,623&4011~~ &3,563.53~~~ &  1,875.18~~~ &0.00097&$-$3.0125 $\pm$ 0.0689\\
2 & 45~~~~ &60~~~~ &4,011~~ &621,371&4006~~ &4,631.72~~~   & 6,141.11~~~ 	&0.00644&$-$2.1906 $\pm$ 0.0437\\
3 & 60~~~~ &75~~~~ &4,006~~ &192,534&4028~~ &5,719.86~~~   & 13,329.35~~~ &0.02092&$-$1.6794 $\pm$ 0.0330\\
4 & 75~~~~ &95~~~~ &4,028~~ &146,702&3017~~ &9,388.65~~~   & 22,684.85~~~ &0.02056&$-$1.6868 $\pm$ 0.0322\\
5 & 95~~~~ &120~~~~ &3,017~~ &124,196&3015~~ &	14,108.38~~~   & 36,705.21~~~ 	&0.02427&$-$1.6148 $\pm$ 0.0305\\
6 & 120~~~~ &155~~~~ &3,015~~ &117,846&2613~~ &25,059.22~~~  & 56,659.00~~~ &0.02217&$-$1.6541 $\pm$ 0.0292\\
7 & 155~~~~ &195~~~~ &2,613~~ &44,686&2511~~ &	35,535.68~~~   & 90,545.96~~~ &0.05619&$-$1.2503 $\pm$ 0.0275\\
8 & 195~~~~ &245~~~~ &2,511~~ &19,897&2513~~ &	58,703.24~~~  &  140,922.01~~~ &0.12630&$-$0.8985 $\pm$ 0.0253\\
9 & 245~~~~ &305~~~~ &2,513~~ &7,236&2532~~ &92,995.11~~~  &  225,326.57~~~ 	&0.34991&$-$0.4560 $\pm$ 0.0219\\
10 & 305~~~~ &385~~~~ &2,532~~ &3,302&2536~~ &126,023.89~~~   & 325,272.66~~~ &0.76801&$-$0.1146 $\pm$ 0.0126\\
[1ex]
\hline
\end{tabular}
\label{table:nonlin}
\end{table*}

\begin{table*}[ht]
\caption{Technical specifications of jFFS iterations for $L=14.0226$~nm. $N_c, N_t$ and $N_s$ correspond to the number of starting configurations, trial trajectories and number of successful crossings, respectively. $t_{\text{succ}}$ and $t_{\text{fail}}$ are the average length of a successful and a failing trajectory, respectively. }
\centering
\begin{tabular}{c r r r r r r r c c}
\hline\hline
$k$ & ~~~$\lambda_k$~~~ & $\lambda_{k+1}$~~ & $N_c$~~ & $N_t$~~~~ & ~~~$N_s$~~~ & $t_\text{succ}$~[fs]~~ & $t_\text{fail}$~[fs]~~ &$P(\lambda_{k+1}|\lambda_k)$ & $\log_{10}\left[P(\lambda_{k+1}|\lambda_k)\right]$  \\ [0.5ex] 
\hline
0 & 15~~~~ &27~~~~ &16,975~~ &1,605,092&4,046~~ &1,844.33~~~  &  1,049.76~~~ 		&0.00252&$-$2.5984 $\pm$ 0.0343\\
1 & 27~~~~ &42~~~~ &4,046~~ &1,787,419&4,016~~ &	3,397.13~~~  &  2,742.75~~~ 	&0.00224&$-$2.6484 $\pm$ 0.0668\\
2 & 42~~~~ &56~~~~ &4,016~~ &423,667&4,026~~ & 3,956.49~~~   & 6,903.61~~~ 	&	0.00950&$-$2.0221 $\pm$ 0.0418\\
3 & 56~~~~ &72~~~~ &4,026~~ &278,980&3,520~~ &5,673.30~~~  &  12,890.44~~~ 	&	0.01261&$-$1.8990 $\pm$ 0.0368\\
4 & 72~~~~ &88~~~~ &3,520~~ &118,800&3,521~~ &7,288.23~~~   & 21,622.35~~~ 	&	0.02963&$-$1.5281 $\pm$ 0.0311\\
5 & 88~~~~ &106~~~~ &3,521~~ &67,083&3,025~~ & 9,451.11~~~   & 32,747.28~~~ &	0.04509&$-$1.3458 $\pm$ 0.0290\\
6 & 106~~~~ &126~~~~ &3,025~~ &35,841&2,516~~ &	12,144.92~~~  & 46,185.32~~~  &	0.07019&$-$1.1536 $\pm$ 0.0283\\
7 & 126~~~~ &150~~~~ &2,516~~ &31,563&2,511~~ &	16,293.37~~~   & 62,393.97~~~  &	0.07955&$-$1.0993 $\pm$ 0.0281\\
8 & 150~~~~ &185~~~~ &2,511~~ &38,246&2,502~~ &	 29,007.67~~~  &  86,733.23~~~  &	0.06541&$-$1.1842 $\pm$ 0.0276\\
9 & 185~~~~ &225~~~~ &2,502~~ &18,333&2,517~~ &	42,548.65~~~  &  127,164.90~~~  &	0.13729&$-$0.8623 $\pm$ 0.0255\\
10 & 225~~~~ &275~~~~ &2,517~~ &7,714&2,002~~ &	 70,453.57~~~   & 186,720.91~~~  &	0.25952&$-$0.5858 $\pm$ 0.0264\\
11 & 275~~~~ &335~~~~ &2,002~~ &4,425&2,517~~ &	100,555.14~~~  &  279,670.14~~~  &	0.56881&$-$0.2450 $\pm$ 0.0177\\
12 & 335~~~~ &410~~~~ &2,517~~ &3,352&3,016~~ &	111,917.75~~~   & 364,631.98~~~  &	0.89976&$-$0.0459 $\pm$ 0.0076\\[1ex]
\hline
\end{tabular}
\label{table:nonlin}
\end{table*}

\begin{table*}
\caption{Technical specifications of jFFS iterations for $L=15.9348$~nm. $N_c, N_t$ and $N_s$ correspond to the number of starting configurations, trial trajectories and number of successful crossings, respectively. $t_{\text{succ}}$ and $t_{\text{fail}}$ are the average length of a successful and a failing trajectory, respectively. }
\centering
\begin{tabular}{c r r r r r r r c c}
\hline\hline
$k$ & ~~~$\lambda_k$~~~ & $\lambda_{k+1}$~~ & $N_c$~~ & $N_t$~~~~ & ~~~$N_s$~~~ & $t_\text{succ}$~[fs]~~ & $t_\text{fail}$~[fs]~~ &$P(\lambda_{k+1}|\lambda_k)$ & $\log_{10}\left[P(\lambda_{k+1}|\lambda_k)\right]$  \\ [0.5ex] 
\hline
0 & 15~~~~ &28~~~~ &20,275~~ &1528393&3845~~ &3,219.38~~~  &  2,184.64~~~ 	&0.00251&$-$2.5993 $\pm$ 0.0304 \\
1 & 28~~~~ &38~~~~ &3,845~~ &201,230&3,001~~ &2,431.01~~~   & 4,750.95~~~ 	&0.01491&$-$1.8264 $\pm$ 0.0413\\
2 & 38~~~~ &52~~~~ &3,001~~ &417,330&3,187~~ &4,091.60~~~  & 8,139.72~~~  &0.00763&$-$2.1171 $\pm$ 0.0470\\
3 & 52~~~~ &65~~~~ &3,187~~ &123,698&2,634~~ &4,451.55~~~  & 13,727.40~~~ &0.02129&$-$1.6717 $\pm$ 0.0404\\
4 & 65~~~~ &79~~~~ &2,634~~ &73,182&2,612~~ &5,731.73~~~  & 20,689.06~~~  &0.03569&$-$1.4474 $\pm$ 0.0350\\
5 & 79~~~~ &93~~~~ &2,612~~ &37,699&2,539~~ & 6,280.37~~~  & 28,670.39~~~ &0.06734&$-$1.1717 $\pm$ 0.0318\\
6 & 93~~~~ &111~~~~ &2,539~~ &42,724&2,509~~ &9,480.07~~~   & 38,638.63~~~ 		&0.05872&$-$1.2312 $\pm$ 0.0310\\
7 & 111~~~~ &128~~~~ &2,509~~ &21,619&2,545~~ &10,210.76~~~  &  51,340.52~~~ 	&0.11772&$-$0.9291 $\pm$ 0.0279\\
8 & 128~~~~ &148~~~~ &2,545~~ &20,971&2,654~~ &	14,118.70~~~  & 65,593.54~~~  &0.12655&$-$0.8977 $\pm$ 0.0269\\
9 & 148~~~~ &168~~~~ &2,654~~ &12,880&2,513~~ &15,158.73~~~   & 83,074.95~~~  &0.19510&$-$0.7097 $\pm$ 0.0258\\
10 & 168~~~~ &188~~~~ &2,513~~ &8,552&2,538~~ &	17,699.95~~~  & 102,810.05~~~ 	&0.29677&$-$0.5275 $\pm$ 0.0233\\
11 & 188~~~~ &215~~~~ &2,538~~ &11,602&3,182~~ &	29,056.92~~~  & 128,895.05~~~ &0.27426&$-$0.5618 $\pm$ 0.0210\\
12 & 215~~~~ &245~~~~ &3,182~~ &6,831&2,517~~ &35,909.36~~~  &  164,267.26~~~ 	&0.36846&$-$0.4336 $\pm$ 0.0213\\
13 & 245~~~~ &285~~~~ &2,517~~ &5,718&2,632~~ &	57,148.41~~~  & 216,271.88~~~ &0.46030&$-$0.3370 $\pm$ 0.0194\\
14 & 285~~~~ &340~~~~ &2,632~~ &4,809&3,215~~ &	93,943.50~~~   & 305,798.31~~~ 	&0.66853&$-$0.1749 $\pm$ 0.0137\\
15 & 340~~~~ &410~~~~ &3,215~~ &2,729&2,513~~ &108,253.04~~~  &  394,034.58~~~ 	&0.92085&$-$0.0358 $\pm$ 0.0071\\[1ex]
\hline
\end{tabular}
\label{table:nonlin}
\end{table*}

\begin{table*}
\caption{Technical specifications of jFFS iterations for $L=17.8470$~nm. $N_c, N_t$ and $N_s$ correspond to the number of starting configurations, trial trajectories and number of successful crossings, respectively. $t_{\text{succ}}$ and $t_{\text{fail}}$ are the average length of a successful and a failing trajectory, respectively. }
\centering
\begin{tabular}{c r r r r r r r c c}
\hline\hline
$k$ & ~~~$\lambda_k$~~~ & $\lambda_{k+1}$~~ & $N_c$~~ & $N_t$~~~~ & ~~~$N_s$~~~ & $t_\text{succ}$~[fs]~~ & $t_\text{fail}$~[fs]~~ &$P(\lambda_{k+1}|\lambda_k)$ & $\log_{10}\left[P(\lambda_{k+1}|\lambda_k)\right]$  \\ [0.5ex] 
\hline
0 & 15~~~~ &28~~~~ &21,798~~ &1,009,164&4,077~~ &5,560.00~~~  & 5,277.67~~~ 	&0.00403& $-$2.3936 $\pm$ 0.0240 \\
1 & 28~~~~ &39~~~~ &4,077~~ &387,112&4,426~~ &	2,926.52~~~  & 8,878.86~~~ 	&0.01143&$-$1.9418 $\pm$ 0.0426\\
2 & 39~~~~ &56~~~~ &4,426~~ &478,497&2,630~~ &	4,628.21~~~  &13,096.29~~~ 	&0.00549&$-$2.2599 $\pm$ 0.0507\\
3 & 56~~~~ &70~~~~ &2,630~~ &121,420&2,583~~ &	4,655.45~~~  & 20,407.38~~~ 	&0.02127&$-$1.6721 $\pm$ 0.0399 \\
4 & 70~~~~ &85~~~~ &2,583~~ &75,157&2,534~~ &	6,345.58~~~  & 27,819.77~~~ 	&0.03371&$-$1.4722 $\pm$ 0.0378\\
5 & 85~~~~ &100~~~~ &2,534~~ &45,776&3,286~~ &	7,352.74~~~  & 37,859.79~~~ 	&0.07178&$-$1.1440 $\pm$ 0.0293\\ 
6 & 100~~~~ &120~~~~ &3,286~~ &43,726&2,536~~ &	11,279.17~~~  & 49,349.58~~~ &0.05799&$-$1.2366 $\pm$ 0.0291\\ 
7 & 120~~~~ &140~~~~ &2,536~~ &24,591&2,503~~ &13,063.74~~~   &  64,884.04~~~ 	&0.10178&$-$0.9923 $\pm$ 0.0281\\
8 & 140~~~~ &160~~~~ &2,503~~ &14,577&2,506~~ &15,602.43~~~  &   82,774.09~~~  &0.17191&$-$0.7647 $\pm$ 0.0260\\
9 & 160~~~~ &188~~~~ &2,506~~ &17,288&2,532~~ &	25,312.36~~~  & 105,096.12~~~ &0.14645&$-$0.8343 $\pm$ 0.0260\\
10 & 188~~~~ &220~~~~ &2,532~~ &11,764&2,517~~ &	35,904.43~~~  &   138,730.74~~~ 	&0.21395&$-$0.6696 $\pm$ 0.0247\\
11 & 220~~~~ &260~~~~ &2,517~~ &8,206&2,528~~ &	54,797.87~~~  & 188,253.00~~~ &0.30806& $-$0.5113 $\pm$ 0.0226\\
12 & 260~~~~ &330~~~~ &2,528~~ &5,747&2,516~~ &121,431.38~~~  &   275,286.59~~~ 		&0.43779&$-$0.3587 $\pm$ 0.0201\\
13 & 330~~~~ &430~~~~ &2,516~~ &2,846&2,515~~ &140,264.78~~~  &  392,421.36~~~ 		&0.88369&$-$0.0537 $\pm$ 0.0087\\
[1ex]
\hline
\end{tabular}
\label{table:nonlin}
\end{table*}

\end{document}